\pdfoutput=1
\documentclass[12pt,a4]{article}
\usepackage{a4wide}
\usepackage[hidelinks]{hyperref}
\usepackage{latexsym}
\usepackage{cite}
\usepackage{mathrsfs}
\usepackage{amssymb}
\usepackage{amsmath}
\usepackage{graphicx}
\usepackage{braket}
\usepackage{enumerate}
\usepackage[usenames,dvipsnames]{xcolor}
\usepackage{todonotes}

\newcommand{\cA}{\mathcal{A}}

\newcommand{\cN}{\mathcal{N}}

\newcommand{\cP}{\mathcal{P}}

\newcommand{\vol}{\mathrm{Vol}}

\newcommand{\GL}{\mathrm{GL}}

\newcommand{\diag}{\, \mathrm{diag}}
\newcommand{\rd}{\, \mathrm{d}}

\newcommand{\be}{\begin{equation}\label}
\newcommand{\ee}{\end{equation}}
\newcommand{\bea}{\begin{eqnarray}\label}
\newcommand{\eea}{\end{eqnarray}}

\date{}
\begin{document}

\title{On-Shell Diagrams for $\mathcal{N}=8$ Supergravity Amplitudes}
\author{Paul Heslop and Arthur E. Lipstein \vspace{7pt}\\ \normalsize \textit{
Department of Mathematical Sciences}\\\normalsize\textit{Durham University, Durham, DH1 3LE, United Kingdom}}
\maketitle
\begin{abstract}
We define recursion relations for $\mathcal{N}=8$ supergravity amplitudes
using a generalization of the on-shell diagrams developed for planar $\mathcal{N}=4$
super-Yang-Mills. Although the recursion relations generically give
rise to non-planar on-shell diagrams, we show that at tree-level
  the recursion can be chosen to yield only  planar diagrams, the same
  diagrams occurring in the planar $\cN=4$ theory. This implies
  non-trivial identities for non-planar diagrams as well as interesting
  relations between the $\mathcal{N}=4$ and $\cN=8$ theories. We show that the on-shell diagrams of $\mathcal{N}=8$ supergravity obey equivalence relations analogous to those of $\mathcal{N}=4$ super-Yang-Mills, and we develop a systematic algorithm for reading off Grassmannian integral formulae directly from the on-shell diagrams. We also show that the 1-loop 4-point amplitude of $\mathcal{N}=8$ supergravity can be obtained from on-shell diagrams.   
\end{abstract}

\pagebreak
\tableofcontents

\section{Introduction}

Standard Feynman diagram techniques often obscure the underlying
simplicity of on-shell scattering amplitudes. One reason for this is
that individual Feynman diagrams are not gauge invariant and contain
unphysical degrees of freedom. This difficulty can be overcome  by
working with on-shell diagrams \cite{ArkaniHamed:2012nw}, which are
built out of 3-point vertices using BCFW recursion
\cite{Britto:2004ap,Britto:2005fq} and do not contain virtual
particles. Moreover, scattering amplitudes often exhibit symmetries
which are hidden from the point of view of the spacetime
Lagrangian. In the case of $\mathcal{N}=4$ super-Yang-Mills (SYM)
\cite{Brink:1976bc}, on-shell diagrams make the Yangian symmetry of
the amplitudes manifest and reveal an underlying Grassmannian structure \cite{ArkaniHamed:2009si,ArkaniHamed:2009dn,ArkaniHamed:2009dg}. 

The Yangian symmetry arises from combining ordinary superconformal
symmetry with dual superconformal symmetry
\cite{Drummond:2007aua,Drummond:2008vq,Brandhuber:2008pf}, which
provides a canonical definition for the loop integrand of the planar
$\mathcal{N}=4$ SYM S-matrix, ultimately making it possible to extend
BCFW recursion to loop-level \cite{ArkaniHamed:2010kv}. BCFW recursion for loop amplitudes was also studied in \cite{Berger:2006ci,CaronHuot:2010zt,Boels:2010nw}. On-shell
diagrams also reveal an underlying cluster algebra structure in
$\mathcal{N}=4$ SYM amplitudes which is encoded in the dlog form of
loop integrands (this form was simultaneously derived using the Wilson loop in twistor space \cite{Lipstein:2012vs,Lipstein:2013xra}). There is also evidence that the dlog form of loop integrands persists in the non-planar sector \cite{Arkani-Hamed:2014via,Bern:2014kca,Bern:2015ple}. Ultimately, on-shell diagrams  and their correspondence to positive cells of the Grassmannian suggest a geometric interpretation of scattering amplitudes as the volume of a new object known as the Amplituhedron \cite{Arkani-Hamed:2013jha,Arkani-Hamed:2013kca,Arkani-Hamed:2014dca}.

An important question is how to generalize these ideas beyond planar $\mathcal{N}=4$ SYM. Although there has been some work on non-planar on-shell diagrams \cite{Arkani-Hamed:2014via,Arkani-Hamed:2014bca,Franco:2015rma,Chen:2015qna,Frassek:2016wlg}, on-shell diagrams for form factors in $\mathcal{N}=4$ SYM \cite{Frassek:2015rka}, and amplitudes in $\mathcal{N}<4$ SYM \cite{Benincasa:2015zna}, on-shell diagrams for gravitational amplitudes have so far not been explored. Since gravity amplitudes are intrinsically non-planar, any new results in this direction may also suggest new techniques for computing non-planar YM amplitudes. In this paper, we take the first steps in this direction by developing an on-shell diagram formalism for $\mathcal{N}=8$ supergavity (SUGRA), which is the natural starting point since it is maximally supersymmetric and its amplitudes also exhibit a great deal of simplicity \cite{ArkaniHamed:2008gz}. 

We develop on-shell diagrams for tree-level amplitudes in $\mathcal{N}=8$ SUGRA using BCFW recursion. Our diagrammatic recursion relation is similar to that of $\mathcal{N}=4$ SYM but has some important differences. For example, the BCFW bridge used to combine lower-point on-shell diagrams is modified with respect to the one in $\mathcal{N}=4$ SYM (we will soon see that this simply amounts to adding a decoration to the BCFW bridge of $\mathcal{N}=4$ SYM). Moreover, since gravity amplitudes are permutation invariant -- and there is no concept of colour ordering -- the on-shell diagrams which arise from the recursion relation will generically be non-planar. Nevertheless, we show that it is possible to restrict the recursion relation to a planar sector of on-shell diagrams, from which the full scattering amplitudes are obtained simply by summing over permutations of the external legs. If one chooses to work outside of the planar sector, this gives rise to remarkable new identities for non-planar on-shell diagrams. The on-shell diagrams of $\mathcal{N}=8$ SUGRA also exhibit equivalence relations analogous to those of $\mathcal{N}=4$ SYM such as square moves and mergers. We also show that on-shell diagrams can be easily computed by assigning variables and arrows to the edges of the diagrams, and reading off expressions directly from the diagrams using a simple set of rules.  

Ultimately, this approach gives rise to new Grassmannian integral
formulae for the scattering amplitudes, which further imply a
  form of positivity in the planar sector from which the amplitudes
can be derived. Grassmannian integral formulae for $\mathcal{N}=8$
supergravity amplitudes have previously been deduced from twistor
string theory \cite{Cachazo:2012pz,He:2012er}, and it would be
interesting to see how they are related to our formulae. Finally, we show that the 1-loop 4-point amplitude of $\mathcal{N}=8$ SUGRA can be obtained from on-shell diagrams, which suggests the possiblity of formulating loop-level BCFW recursion in this theory as well.                

\section{Tree-level Recursion} \label{recursion}

As mentioned in the introduction, the difficulties of Feynman diagrams can be overcome by using BCFW recursion relations to express higher-point on-shell amplitudes in terms of lower-point on-shell amplitudes. In four dimensions, an on-shell momentum for a massless particle can be written in the following bispinor form:
\[
p^{\alpha\dot{\alpha}}=\lambda^{\alpha}\tilde{\lambda}^{\dot{\alpha}},
\]
where $\alpha=0,1$ and $\dot{\alpha}=\dot{0},\dot{1}$ are chiral and antichiral spinor indices. For supersymmetric theories, the particles also have supermomentum:
\[
q^{\alpha a}=\lambda^{\alpha}\eta^{a},
\]
where $\eta$ is Grassman odd and $a=1,...,\mathcal{N}$ and $\mathcal{N}$ denotes the amount of supersymmetry. 

The BCFW recursion relations are natrually encoded by on-shell diagrams, which differ from standard Feynman diagrams in that they do not contain virtual particles. The building blocks for on-shell diagrams are 3-point MHV and anti-MHV amplitudes, which encode the scattering of three gluons or gravitons with helicity $\left\{ --+\right\} $ and $\left\{++-\right\} $, respectively. More generally, $n$-point N$^k$MHV amplitudes encode the scattering of $k+2$ particles of negative helicity and $n-k-2$ particles of positive helicity. The 3-point MHV amplitudes of $\mathcal{N}=8$
SUGRA are essentially the square of their $\mathcal{N}=4$
SYM counterparts and are given by
\begin{align}\label{eq:4}
\mathcal{\mathcal{A}}^\text{MHV}_{3}&=\frac{\delta^{8}\left(\left[12\right]\eta_{3}+\left[23\right]\eta_{1}+\left[31\right]\eta_{2}\right)}{\left[12\right]^{2}\left[23\right]^{2}\left[31\right]^{2}}\delta^{4}\left(\lambda_{1}\tilde{\lambda}_{1}+\lambda_{2}\tilde{\lambda}_{2}+\lambda_{3}\tilde{\lambda}_{3}\right),\,\,\,\lambda_{1}\propto\lambda_{2}\propto\lambda_{3}\notag\\
\mathcal{\mathcal{A}}^{\overline{\text{MHV}}}_{3}&=\frac{\delta^{16}\left(\lambda_{1}\eta_{1}+\lambda_{2}\eta_{2}+\lambda_{3}\eta_{3}\right)}{\left\langle 12\right\rangle ^{2}\left\langle 23\right\rangle ^{2}\left\langle 31\right\rangle ^{2}}\delta^{4}\left(\lambda_{1}\tilde{\lambda}_{1}+\lambda_{2}\tilde{\lambda}_{2}+\lambda_{3}\tilde{\lambda}_{3}\right),\,\,\,\tilde{\lambda}_{1}\propto\tilde{\lambda}_{2}\propto\tilde{\lambda}_{3}.
\end{align}
We denote these building blocks with on-shell diagrams using black and white vertices, respectively:
  \begin{align}\label{eq:16}
\includegraphics{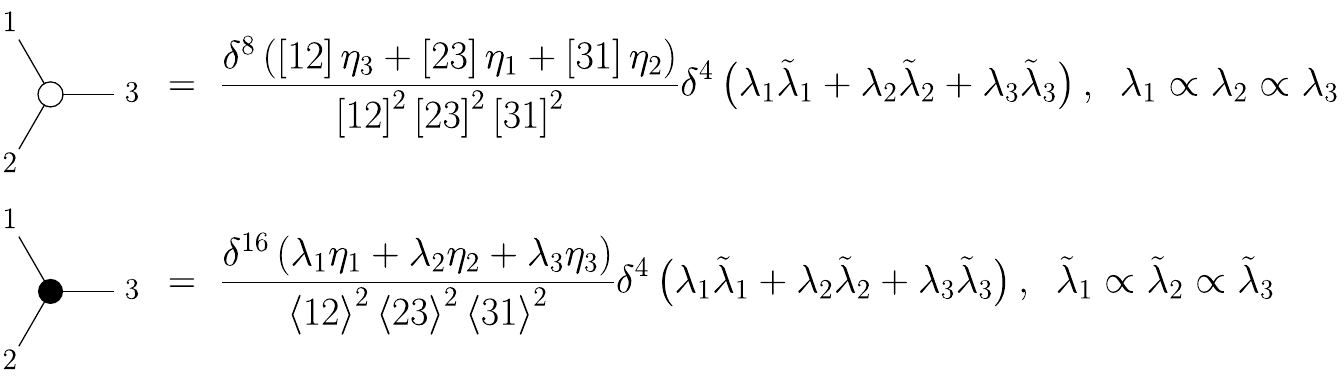}  
  \end{align}
More general on-shell diagrams are constructed by connecting 3-point
vertices and integrating over the on-shell supermomenta
associated with the internal edges between two vertices:
\begin{align}\label{eq:13}
\int d\mu = \int\frac{\rd^{\cN}\eta \rd^{2}\lambda \rd^{2}\tilde{\lambda}}{\vol \,\GL(1)},
\end{align}
where the measure is over $\lambda,\tilde \lambda,\eta$ modulo the
little group phase $\lambda\rightarrow c\lambda,\tilde \lambda
\rightarrow c^{-1}\tilde \lambda$, which we denote by quotienting by
Vol $\mathrm{GL}(1)$.

In order to construct on-shell diagrams corresponding to higher-point amplitudes, one uses the BCFW bridge:
\begin{align}\label{eq:18}
       \includegraphics{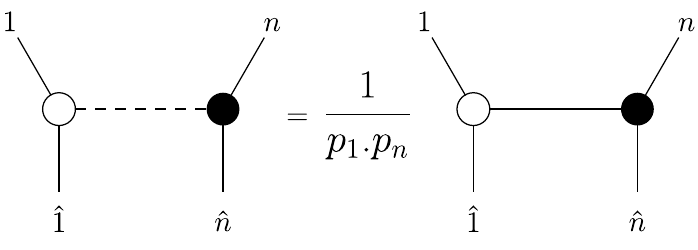}  
\end{align}
 which is essentially a decorated version of the one for $\mathcal{N}=4$ SYM. 
    Parameterizing the momentum through the internal edge 
by $\alpha\lambda_{1}\tilde{\lambda}_{n}$,
one finds that
\[
\lambda_{\hat{1}}\tilde{\lambda}_{\hat{1}}=\lambda_{1}\left(\tilde{\lambda}_{1}-\alpha\tilde{\lambda}_{n}\right)
\]
\[
\lambda_{\hat{n}}\tilde{\lambda}_{\hat{n}}=\left(\lambda_{n}+\alpha\lambda_{1}\right)\tilde{\lambda}_{n}.
\]
Hence, this diagram corresponds to BCFW shifting legs $(1,n)$. In
addition to this, we must multiply the diagram by the factor $1/p_{1} \cdot p_{n}$,
which we indicate by making the central line dashed. Since $p_1 \cdot p_n=\hat p_1 \cdot p_n=p_1 \cdot \hat p_n = \hat p_1 \cdot \hat p_n$, it doesn't matter which two momenta  we choose for the decoration, as long as there is one on either side of the decoration. We will derive this decoration in the next subsection.

\begin{figure}[htbp]
\centering
       \includegraphics{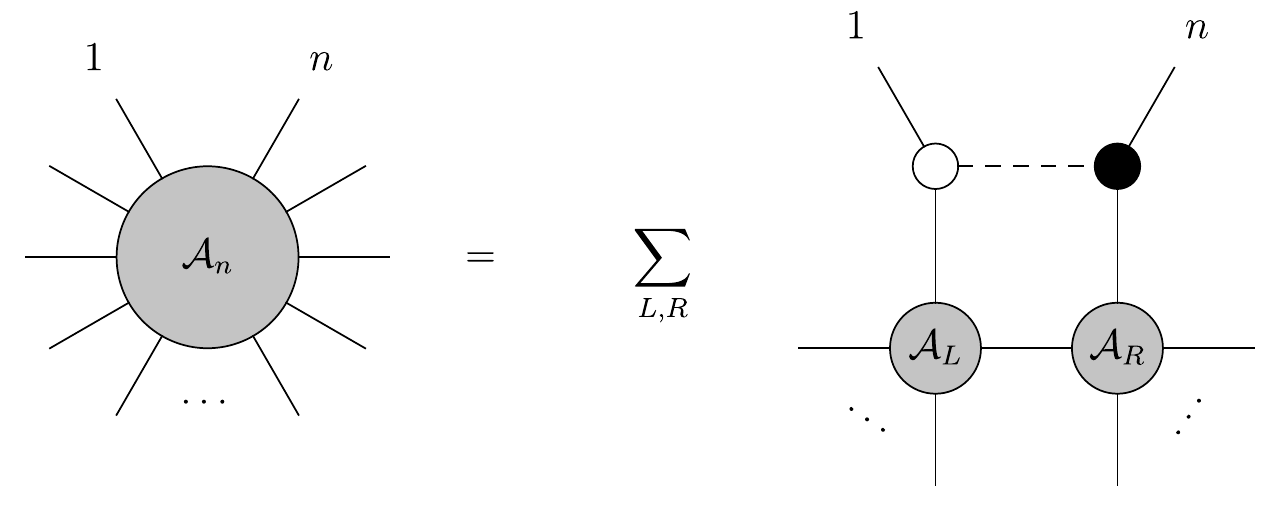}  
    \caption{Construction of amplitudes via recursion using the BCFW
      bridge. For $\mathcal{N}=8$ SUGRA, the  bridge is decorated and the sum is over all ways of partitioning particles $\left\{
        2,...,n-1\right\}$ into two sets $L,R$.}
    \label{tree}
    \end{figure}

Using the above rules, on-shell diagrams for higher-point tree-level scattering
amplitudes can be constructed by connecting on-shell diagrams for
lower-point amplitudes with a BCFW bridge and summing over all permutations
of the unshifted legs, as depicted in Figure~\ref{tree}. 

\subsection{Derivation of the decorated BCFW bridge}

The basic idea underlying BCFW recursion is to deform the momenta of two legs of an on-shell amplitude by a complex parameter. After doing so, the amplitude develops poles in the deformation parameter and the residues correspond to products of lower-point on-shell amplitudes, allowing one to compute higher-point amplitudes from lower-point amplitudes.The BCFW recursion relations can be applied to a very broad range of theories such as Yang-Mills \cite{Britto:2004ap} and gravity \cite{Bedford:2005yy,Cachazo:2005ca} in $d \geq 4$ dimensions, and can also be adapted to $d=3$ \cite{Gang:2010gy}.  The supersymmetric form of the BCFW recursion relation~\cite{Britto:2004ap,Britto:2005fq,Brandhuber:2008pf,ArkaniHamed:2008gz} takes the form
\begin{align}\label{eq:1}
 \cA^{\text{tree}}(\cP) = \sum_{\cP_L(i),\cP_R(j)} \int \frac{d^4pd^\cN\!\! \eta}{p^2}
   \ \cA^{\text{tree}}_L(\hat P, \{\hat \cP(i)\})
  \cA^{\text{tree}}_R(-\hat P,\{\hat \cP(j)\} )\ . 
\end{align}
Here $\cP =\{P_1,\dots,P_n\} $ is
the set of all external supermomenta $P_i=(p_i,\eta_i)$. There are two special
particles, $i,j$  and the sum on the RHS is
over all bipartitions of the particle numbers  such that $i$ is in one partition and
$j$ in the other, with $\cP_L(i)$ and $\cP_R(j)$ the corresponding
sets of supermomenta.
The hats over the external (massless) supermomenta on the RHS indicate  the following
deformations (in spinor helicity form)
\begin{align}
  \hat \lambda_i = \lambda_i + z_p \lambda_j \qquad   \hat {\tilde
  \lambda}_j = \tilde \lambda_j - z_p \tilde \lambda_i
\end{align}
with all other $\lambda_k,\tilde \lambda_k$ remaining
undeformed. Finally the hatted internal supermomenta are defined as
\begin{align}
  \hat p = p-z_p\lambda_j \tilde \lambda_i, 
\end{align}
with $\hat p^2=0$ which fixes
\begin{align}\label{eq:17}
z_p=p^2/(2\langle j|p|i])\ .
\end{align}
A remarkable feature of the BCFW recursion is that the result is
independent of the choice of special points $i,j$. The above formula is valid for both SYM and SUGRA.

We now compare the terms in this BCFW resursion~\eqref{eq:1} with its form as a BCFW bridge.  The idea is that each term in the sum on the RHS of~(\ref{eq:1}) has the interpretation of an on-shell diagram
consisting of two on-shell amplitudes $\cA_L$ and $\cA_R$ (which will themselves can be recursively described via on-shell diagrams) together with a
three-point  MHV
and a three-point $\overline{\text{MHV}}$ amplitude connected with
four internal lines, as depicted in the picture below.
Each internal line yields an integration over
the on-shell supermomentum flowing through it~\eqref{eq:13}.
\begin{center}
  \includegraphics{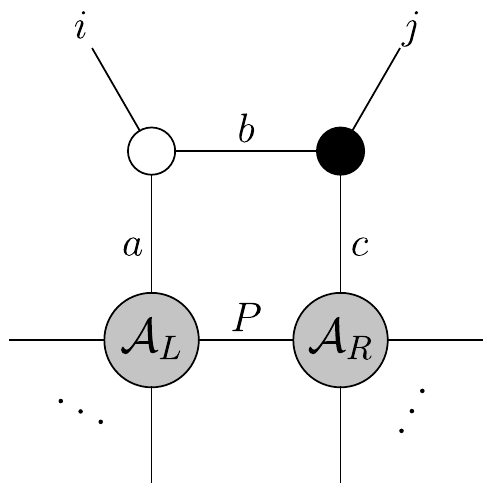}
\end{center}
Hence, this on-shell diagram simply represents
\begin{align}
 \text{diagram}= \int d\mu_ad\mu_bd\mu_cd\mu_P  \cA_3 \bar \cA_3 \cA_L \cA_R \ .
\end{align}
Let us first consider consider integrating the bosonic parts of the measures
$d\mu_ad\mu_bd\mu_c$ against the bosonic delta functions associated
with the 3-point amplitudes. There are nine bosonic integrations and eight bosonic delta functions, so we have one left over integration which when combined with the integral over the on-shell momentum $P$ gives rise to an integral over and off-shell momentum $p$. Writing
\begin{align}
\cA_3(1,2,3) = \delta(p_1+p_2+p_3) A_3(1,2,3)
  \qquad
 \bar \cA_3(1,2,3) = \delta(p_1+p_2+p_3) \bar A_3(1,2,3)
\end{align}
we find that
\begin{align}\label{eq:2}
 \text{diagram}= &\int d\mu_ad\mu_bd\mu_cd\mu_P
  \delta^4(p_i-p_a-p_b)\delta^4(p_j+p_b-p_c) \cA_3
  \bar \cA_3\cA_L \cA_R\notag\\
=&\int  
d^\cN\!\!\eta_ad^\cN\!\!\eta_bd^\cN \!\!\eta_c \int d^\cN\!\!\eta_p \frac{d^4p}{p^2}\times \frac1{p_i.p_j} \times (A_3 \bar A_3 \cA_L \cA_R)|
\end{align}
The $1/p_i \cdot p_j$ arises from Jacobians
and is a key point when considering $\cN=8$ supergravity.
On the RHS, internal momenta have been integrated out against delta
functions and so must
be replaced by the result of this, indicated by the vertical
line. The explicit replacements are
\begin{align}\label{eq:3}
  \lambda_a &= \hat \lambda_i \quad &  \tilde \lambda_a &= \tilde \lambda_i\notag\\
  \lambda_b &= \lambda_j \quad &  \tilde \lambda_b &= -z_p \tilde
                                                          \lambda_i\notag\\
  \lambda_c &= \lambda_j \quad &   {\tilde \lambda_c} &=  {\hat {\tilde \lambda}_j}        \notag\\
p &= \lambda_p \tilde \lambda_p + z_p \lambda_j \tilde \lambda_i
\end{align}
where $z_p$ is defined in~\eqref{eq:17}.
Finally we need to do the integration over internal fermionic degrees
of freedom and consider the explicit form of the three-point
amplitudes. This is where the dependence on $\cN$  comes into the computations
for the first time. The three-point amplitudes are given as:
\begin{align}
  \cA_3(1,2,3)=\frac{\delta^{2\cN}(\eta_1 \lambda_1+\eta_2
  \lambda_2+\eta_3 \lambda_3)
  }{(\langle12\rangle\langle23\rangle\langle31\rangle)^{\cN/4}} \qquad
  \bar \cA_3(1,2,3)=\frac{\delta^{\cN}(\eta_1 [23]+\eta_2
  [31]+\eta_3 [12])
  }{([12][23][31])^{\cN/4}}
\end{align}
where $\mathcal{N}=4,8$ describe $\mathcal{N}=4$ SYM and $\mathcal{N}=8$ SUGRA, respectively.
Integrating $d^\cN\!\!\eta_ad^\cN\!\!\eta_bd^\cN \!\!\eta_c$
against the $3\cN$ fermionic delta functions from the three-point
amplitudes in~(\ref{eq:2}) and inputting the replacements~(\ref{eq:3}), we obtain
\begin{align}
  \text{diagram}=\int d^\cN\!\!\eta_p \frac{d^4p}{p^2}\times
  \frac1{p_i.p_j} \times ( p_i \cdot p_j)^{\cN/4}  \times (\cA_L \cA_R)|\ .
\end{align}
This is the main formula of this section and should be compared with
terms in the
recursion relation~(\ref{eq:1}). We conclude that for $\cN=4$ SYM,
the on-shell diagram precisely corresponds to a term in the BCFW
expansion, but for $\cN=8$ we have an additional power of
$p_i \cdot p_j$ in the numerator. Hence, in both cases we can rephrase BCFW recursion in terms of a sum over on-shell
diagrams, but for $\cN=8$ SUGRA the bridge needs to
be supplemented by an additional $\frac 1{p_i \cdot p_j}$. This is the decoration in~\eqref{eq:18}. Hence, the recursion relation in terms of on-shell diagrams depicted in Figure \ref{tree} holds for any number of legs, since it is equivalent to standard BCFW recursion whose validity for $\mathcal{N}=8$ SUGRA was proven in \cite{ArkaniHamed:2008yf}.

\subsection{Tree-level SUGRA amplitudes from {\em planar} on-shell diagrams}

Since the recursion relation involves summing over permutations of the 
unshifted legs in SUGRA, one generally obtains non-planar on-shell diagrams. However in the recursion relation there are two special adjacent legs which are held fixed. If we always choose these two legs to be the ones which we insert into the recursion relation to obtain higher-point amplitudes, the result will always be a sum of planar  planar graphs. This can be proved via a simple induction argument. Assume that all $n'$-point amplitudes  for $n'<n$ can be expressed  as a sum over planar on-shell diagrams with two fixed adjacent external momenta. Then an $n$-point amplitude can be obtained via the recursion relation (as in figure~\ref{tree}), by inserting these lower-point diagrams into  a larger one. By insisting that we always use the fixed adjacent legs in each subdiagram as the ones we attach to either the bridge or the other subdiagram, we obtain the $n$-point amplitude as a sum over planar on-shell diagrams with two fixed adjacent external momenta and we have completed the induction argument. The structure is illustrated in the following picture which then repeats recursively:

\begin{figure}[htbp]
\centering
       \includegraphics{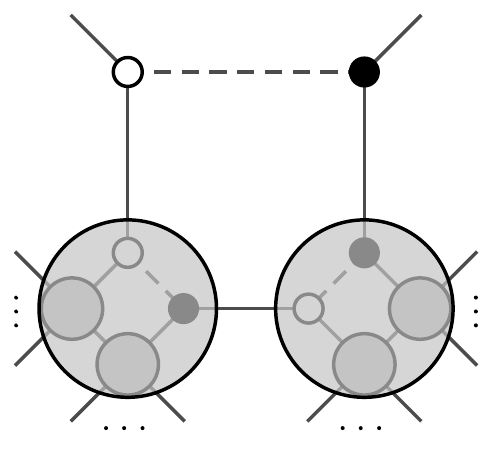}
    \caption{BCFW recursion in terms of planar on-shell diagrams.}
    \label{planar}
    \end{figure}

Thus any tree-level scattering amplitude can be obtained
by summing {\em planar} on-shell diagrams over permutations of unshifted
external legs.  In this way, the amplitudes of tree-level $\mathcal{N}=8$
SUGRA can be associated with planar on-shell diagrams. Indeed the diagrams are precisely the same as those which would appear in $\mathcal{N}=4$ SYM  by recursing in a similar way. The main difference is that in SUGRA we sum over all permutations of the unfixed legs. A structure very reminiscent of this relation between  tree-level SUGRA and SYM was found previously in~\cite{Drummond:2009ge}. It is then interesting to examine what other properties of  planar $\mathcal{N}=4$ amplitudes such as a Grassmannian representation and positivity can be generalized to $\mathcal{N}=8$ supergravity. We will consider this in the next section. 

As an example, consider the 5-point MHV amplitude. If we restrict the recursion relation to a planar sector as described above, the result is given by a sum over six planar diagrams:
\[
       \includegraphics{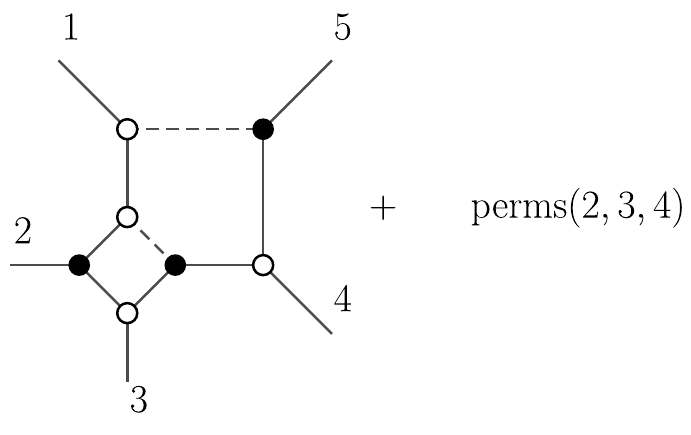}  
     \]
On the other hand, if we apply a recursion in a different way, we will generically get a sum of non-planar diagrams:
\[     
       \includegraphics{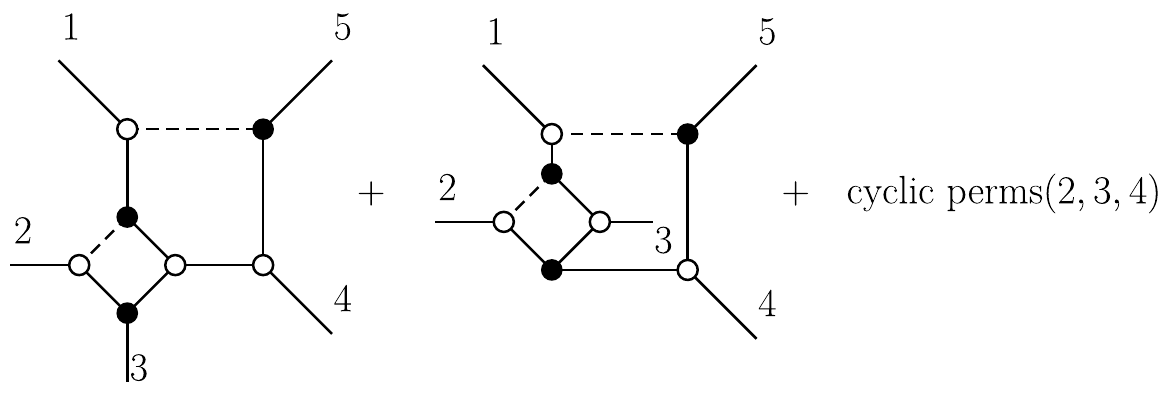}  
     \]
This implies nontrivial relations for non-planar diagrams of $\mathcal{N}=8$ SUGRA. For example, it is straightforward to check the equivalence of the diagrams in the above two Figures using the techniques we describe in the next section. 
     
\section{Grassmannian Representation} \label{grassmannian}

In the previous section, we developed a recursion relation for
tree-level $\mathcal{N}=8$ SUGRA amplitudes in terms of on-shell
diagrams. In this section, we will describe a systematic method for
evaluating the on-shell diagrams, closely following similar
  methods in $\cN=4$ developed in~\cite{ArkaniHamed:2012nw}. In particular we will develop an algorithm for reading off formulae directly from the diagrams in the form of integrals over $k$-dimensional planes in $n$-dimensions, where $n$ and $k$ are the number of external legs and MHV degree, respectively. The space of $k$ planes in $n$ dimensions is known as the Grassmannian $\mathrm{Gr}(k,n)$, which also plays a prominent role in the scattering amplitudes of $\mathcal{N}=4$ SYM. 

Our strategy will be to first write the 3-point amplitudes as Grassmannian integrals and make a special choice of coordinates on the Grassmannian which allow us to read off the integrands directly from the on-shell diagrams by assigning variables and arrows to the edges of the diagram. We then generalize these expressions to higher-point on-shell diagrams by gluing together 3-point vertices and deduce an algorithm for writing down formulae for general on-shell diagrams in terms of their edge variables, which can ultimately be lifted to covariant Grassmannian integral formulae. 

\subsection{3-point amplitudes} \label{3ptg}

 The Grassmanian $\mathrm{Gr}(k,n)$ can be thought of as the set of  $k\times n
 $ matrices $C_{ai}$ modulo the left-action of $\mathrm{GL}(k)$
 \begin{align}
\mathrm{Gr}(k,n)=\left\{ C_{ai}\in M^{k\times n}: C_{ai} \sim g_a{}^b C_{bi}, \quad \forall \ g_a^b \in
 GL(k)\right\}\ .
 \end{align}
Equivalently this is the set of $k$-planes through the
 origin in $n$ dimensions, with the $\mathrm{GL}(k)$ equivalence simply corresponding to
 the freedom of the choice of basis for the $k$-plane. We can then define $C^\perp$
 is the orthogonal $n-k$  plane whose minors ($i j \dots
 k)^\perp= C^\perp_{ai} C^\perp_{bj}\dots C^\perp_{ck}\epsilon^{ab\dots c}$ are determined in terms
 of the minors of $C$ ($i j \dots
 k)= C_{ai} C_{bj}\dots C_{ck}\epsilon^{ab\dots c}$ via
 \begin{align}
   (i_{k+1} \dots i_n)^\perp = (i_1i_2 \dots i_k)\epsilon^{i_1i_2 \dots i_k}{}_{i_{k+1} \dots 
   i_n}\ .
 \end{align}
The natural Grassmanian invariant measure can be written explicitly as
\begin{align}
  \frac{d^{k\times n}C}{GL(k)} = (i_1\dots i_k)dC_{1i_{k+1}}
  \dots dC_{1i_n} \epsilon_{i_1\dots i_n}  \dots  (j_1\dots j_k) dC_{kj_{k+1}}
  \dots dC_{kj_n} \epsilon_{j_1\dots j_n}\ . 
\end{align}
With this measure one can choose any $k(n-k)$ independent coordinates
for $\mathrm{Gr}(k,n)$ and simply plug into the above $k(n-k)$-form.

 One can write the 3-point MHV amplitude  in supergravity as an integral over the Grassmannian
$\mathrm{Gr}(2,3)$ as follows:
\begin{equation}
\mathcal{A}_{3}(1,2,3)=\int\frac{d^{2\times3}C}{GL(2)}\frac{\delta^{4|16}\left(C_{a}\cdot\tilde{\lambda}|C_{a}\cdot\tilde{\eta}\right)\delta^{2}\left(\lambda\cdot C^{\perp}\right)}{\left(12\right)^{2}\left(23\right)^{2}\left(31\right)^{2}}\frac{\left\langle ij\right\rangle }{\left(ij\right)},\label{eq:3ptgrass}
\end{equation}
where $i,j$ are any pair of external legs and $(ij)$ is the minor obtained from columns $i$ and $j$ of the $2 \times 3$ $C$-matrix. 
Note that the delta functions (which imply that $\lambda_i$ is
perpendicular to $C^\perp$ and hence parallel to $C$)
imply that $C_{a i}=H_{ab}\lambda_{i}^{b}$, where
$H\in \mathrm{GL}(2)$. It follows that 
\begin{equation}
\frac{\left\langle ij\right\rangle }{\left(ij\right)}=\det H\label{eq:an}
\end{equation}
 so this ratio is the same for any pair of legs $i,j$.  One can verify directly that this expression is
$\mathrm{GL}(2)$ invariant, permutation invariant (thanks to~(\ref{eq:an})), and
gives the correct result~(\ref{eq:4}) on making any coordinate choice
for the Grassmannian, which we will see shortly. As we show in Appendix \ref{linka}, \eqref{eq:3ptgrass} can also be derived by Fourier transforming the 3-point MHV amplitude in \eqref{eq:4} to twistor space, which gives rise to a ``link representation"  and makes manifest the fact that the amplitude does not have conformal symmetry, since the angle bracket $\left\langle ij\right\rangle$ in \eqref{eq:3ptgrass} is expressed in terms of an infinity twistor. 

 Performing similar manipulations,
we obtain the following Grassmannian $\mathrm{Gr}(1,3)$ integral formula for the 3-point anti-MHV amplitude:
\begin{equation}
\bar \cA_{3}=\int\frac{d^{1\times3}C}{GL(1)}\frac{\delta^{2|8}\left(C\cdot\tilde{\lambda}|C\cdot\tilde{\eta}\right)\delta^{4}\left(\lambda\cdot C^{\perp}\right)}{\left(1\right)^{2}\left(2\right)^{2}\left(3\right)^{2}}\frac{\left[ij\right]}{\left(ij\right)^{\perp}}\label{eq:3ptmb}
\end{equation}
where $i,j$ are once again any pair of external legs and $\left(ij\right)^{\perp}$ is the minor obtained from rows $i$ and $j$ of the $3 \times 2$ $C^\perp$ matrix. In this case,
the delta functions imply that $C_{ia}^{\perp}=\tilde{\lambda}_{i}^{b}G_{ba}$,
where $G\in GL(2)$, so
\begin{equation}
\frac{\left[ij\right]}{(ij)^{\perp}}=\det G\label{eq:sq}
\end{equation}
so this ratio is the same for any pair of external legs.

\subsection{Edge variables and perfect orientations}

When gluing together 3-point amplitudes to form higher-point on-shell diagrams it
is useful to make a particular coordinate choice for the 3-point Grassmanians which allows us to interpret the integrands in terms of
``edge variables'' and provides a systematic way to write down formulae for higher-point on-shell diagrams~\cite{ArkaniHamed:2012nw}.

For the black (MHV) vertex we choose coordinates
\begin{align}
  C_\text{MHV}=\left(
  \begin{array}{ccc}
    1&0&-\alpha_1\alpha_3\\
    0&1&-\alpha_2\alpha_3
  \end{array}
\right)
\qquad \qquad
C^\perp_\text{MHV}=\left(
  \begin{array}{ccc}
   - \alpha_1\alpha_3&-\alpha_2\alpha_3&1
  \end{array}
\right)\ .
\end{align}
whereas for the white (${\overline{\text{MHV}}}$) vertices we choose
\begin{align}
  C_{\overline{\text{MHV}}}=\left(
  \begin{array}{ccc}
    -\alpha_1\alpha_3&-\alpha_2\alpha_3&1
  \end{array}
\right)
\qquad \qquad
C^\perp_{\overline{\text{MHV}}}=\left(
  \begin{array}{ccc}
    1&0&-\alpha_1\alpha_3\\
    0&1&-\alpha_2\alpha_3
  \end{array}
\right)\ .
\end{align}
and we display this choice of coordinates via arrows as:
\begin{align}
  \includegraphics{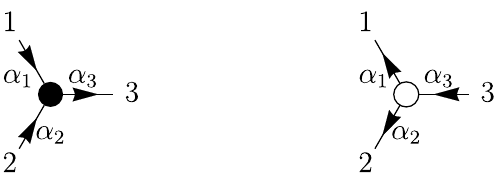}
\end{align}
The relations among the $\lambda$'s and $\tilde \lambda$'s implied by
the delta functions in \eqref{eq:3ptgrass} and \eqref{eq:3ptmb} can now be read off
directly by following paths in the oriented diagrams. For the black node, the delta functions with this choice of $C$ imply that 
$\tilde\lambda_1=\alpha_1\alpha_3\tilde{\lambda}_3$ and
$\tilde\lambda_2=\alpha_2\alpha_3\tilde{\lambda}_3$, whereas for the
white node we have 
$\tilde\lambda_3=\alpha_1\alpha_3\tilde{\lambda}_1+\alpha_2\alpha_3\tilde{\lambda}_2$.%
\footnote{The reader may notice that $C^\perp$ is
  not in fact perpendicular to $C$ (in the  Euclidean sense) with this choice. Indeed we choose momentum flow to follow the arrows, thus momentum conservation reads $p_1+p_2-p_3= \lambda_i \eta^{ij} \tilde \lambda_j= 0$ for the black node, leading to the non-Euclidean metric $\eta^{ij}=\diag(+,+,-)$. $C$ and $C^\perp$ are orthogonal with respect to this metric.} These equations relate a $\tilde \lambda$ associated with an ingoing arrow to $\tilde \lambda$'s
associated with outgoing arrows by summing over all  paths originating from the ingoing arrow in question:  
\begin{align}\label{eq:5}
  \tilde \lambda_i=\sum_{\substack{\text{paths}\\ i\rightarrow j}}
  \Big(\prod_{\substack{\text{edges}\\\text{in path}: e}}\alpha_e
  \Big)\tilde \lambda_j\ .
\end{align}
and similar relations hold for the $\eta$'s.  Similarly, the relations among the $\lambda$'s which arise from the delta functions involving $C^\perp$ arise from summing over the reverse paths: 
\begin{align}\label{eq:6}
  \lambda_i=\sum_{\substack{\text{paths}\\ i\leftarrow j}}
  \Big(\prod_{\substack{\text{edges}\\\text{in path}: e}}\alpha_e
  \Big)
\lambda_j\ .
\end{align}
We can thus read off $C$ and $C^\perp$ directly from the arrows in the on-shell diagrams. 

With the above choices for $C$ and $C^\perp$, the Grassmannian
formulae~(\ref{eq:3ptgrass}) and (\ref{eq:3ptmb}) become
\begin{align}
  \mathcal{A}_{3}={\langle
  12\rangle }\int d(\alpha_1\alpha_3)d(\alpha_2\alpha_3)
 \frac{ \prod_{i=1}^2\delta^{2|8}\left( (\tilde
  \lambda_i|\eta_i)-\alpha_i\alpha_3(\tilde
  \lambda_3|\eta_3)\right)\delta^{2}\left(\lambda_3+\alpha_1\alpha_3\lambda_1+\alpha_2\alpha_3\lambda_2\right)}{\alpha_1^2\alpha_2^2\alpha_3^4}\ .
\label{3alpha}
\end{align}
\begin{align}
  \bar{\mathcal{A}}_{3}={[
  12] }\int d(\alpha_1\alpha_3)d(\alpha_2\alpha_3)
 \frac{
  \delta^{2|8}\left((\tilde \lambda_3|\eta_3)-\alpha_1\alpha_3(\tilde
  \lambda_1|\eta_1)-\alpha_2\alpha_3(\tilde \lambda_2|\eta_2)\right)\prod_{i=1}^2\delta^{2}\left(
  \lambda_i+\alpha_i\alpha_3\lambda_3\right)}{\alpha_1^2\alpha_2^2\alpha_3^4}\ .
\label{3beta}
\end{align}
We can easily recover the original expressions for the
amplitudes in \eqref{eq:4}. For example, if we choose $\alpha_1,\alpha_2$ as our integration variables and integrate them against the final delta function in \eqref{3alpha}, this gives $\alpha_1 = \langle23\rangle/\alpha_3\langle12\rangle$ and $\alpha_2 =
\langle31\rangle/\alpha_3\langle12\rangle$ along with the
Jacobian factor $1/\alpha_3^2\langle12\rangle$. One finds that
$\alpha_3$ drops out and using
\begin{align}
  \delta^{2|8}\Big((\tilde \lambda_1|\eta_1)-\langle 23\rangle/\langle12\rangle(\tilde
  \lambda_3|\eta_3)\Big)\delta^{2|8}\Big((\tilde \lambda_2|\eta_2)-\langle 31\rangle/\langle12\rangle(\tilde
  \lambda_3|\eta_3)\Big)=\langle12\rangle^{-6}\delta^{4|16}\Big(\sum_i(\lambda_i\tilde \lambda_i|\lambda_i\eta_i)\Big)\end{align}
one indeed obtains the 3-point MHV amplitude in \eqref{eq:4}.  

These formulae can be generalized to any on-shell diagram. In particular,
one can always put arrows on an on-shell graph such that
each white node has one incoming and two outgoing arrows, and each
black node has two incoming and one outgoing, known as a ``perfect
orientation''~\cite{ArkaniHamed:2012nw}. One can then associate $\alpha$'s with edges of the graph and read off a formula for the graph in terms of these edge variables which can then be lifted to a Grassmannian integral formula. We will illustrate this for a few simple examples and then spell out a general algorithm. 

\subsection{On-shell diagrams with two vertices} \label{2v}

Let us consider the next simplest examples, notably on-shell diagrams involving two 3-point vertices. Working out these examples in detail will help us deduce an algorithm for evaluating general on-shell diagrams. First consider a two-node diagram in which the vertices have
the same color:
\[  \mathcal{A}_{bbs}=    \raisebox{-.5\height}{\includegraphics{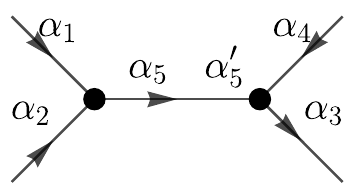}}  
\]
In $\mathcal{N}=4$ SYM, such diagrams obey certain identities which
allow one to define a four-point vertex from merging two three-point vertices.
We will derive analogous identities for $\mathcal{N}=8$ SUGRA. 

Using the formulae for 3-point vertices in \eqref{3alpha} and \eqref{3beta}, this diagram is given by
\begin{equation}
  \mathcal{A}_{bbs}=
\int\frac{d^{2}\lambda_{5}d^{2}\tilde{\lambda}_{5}d^{8}\eta_{5}}{GL(1)}\mathcal{A}_{3}^{L}\mathcal{A}_{3}^{R}\label{eq:bb}
\end{equation}
where
\[
\mathcal{A}_{3}^{L}=\int\frac{d(\alpha_{1}\alpha_5)d(\alpha_{2}\alpha_5)}{\alpha_{1}^{2}\alpha_{2}^{2}\alpha_{5}^{4}}\langle12\rangle\delta^{2|8}\left(\tilde{\lambda}_{1}-\alpha_{1}\alpha_{5}\tilde{\lambda}_{5}\right)\delta^{2|8}\left(\tilde{\lambda}_{2}-\alpha_{2}\alpha_{5}\tilde{\lambda}_{5}\right)\delta^{2}\left(\lambda_{5}-\alpha_{5}\left(\alpha_{1}\lambda_{1}+\alpha_{2}\lambda_{2}\right)\right)
\]

\[
\mathcal{A}_{3}^{R}=\int\frac{d(\alpha_4\alpha_{3})d(\alpha_{5'}\alpha_3)}{\alpha_{3}^{4} \alpha_{4}^{2}\alpha_{5'}^{2}}\left\langle 54\right\rangle \delta^{2|8}\left(\tilde{\lambda}_{5'}-\alpha_{3}\alpha_{5'}\tilde{\lambda}_{3}\right)\delta^{2|8}\left(\tilde{\lambda}_{4}-\alpha_{3}\alpha_{4}\tilde{\lambda}_{3}\right)\delta^{2}\left(\lambda_{3}-\alpha_{3}\left(\alpha_{5'}\lambda_{5'}+\alpha_{4}\lambda_{4}\right)\right)
\]
and we define $\lambda_{5'}=\lambda_{5}$, $\tilde{\lambda}_{5'}=\tilde{\lambda}_{5}$.
Note that a factor of $\alpha_{5}^{4}$ appears in $\mathcal{A}_{3}^{L}$
because it is associated with an outgoing line on a black vertex. Although we can fix one edge variable for each vertex,
we will keep them all unfixed for now in order to be as general as possible. Noting that
\[
\left\langle 54\right\rangle =\frac{\left\langle
    34\right\rangle}{\alpha'_5 \alpha_3} 
\]
\eqref{eq:bb} becomes 
\begin{align*}
  \mathcal{A}_{bbs}=&
\left\langle 12\right\rangle \left\langle 34\right\rangle
                     \int\frac{d(\alpha_{1}\alpha_5)d(\alpha_{2}\alpha_5)d(\alpha_{4}\alpha_3)d(\alpha_{5'}\alpha_3)}{\alpha_{1}^{2}\alpha_{2}^{2}\alpha_{3}^{5}\alpha_{4}^{2}\alpha_{5}^{4}\alpha_{5'}^3}\frac{d^{2}\lambda_{5}d^{2}\tilde{\lambda}_{5}d^{8}\eta_5}{GL(1)}
\\&\times\delta^{2|8}\left(\tilde{\lambda}_{5}-\alpha_{3}\alpha_{5'}\tilde{\lambda}_{3}\right)\delta^{2}\left(\lambda_{5}-\alpha_{5}\left(\alpha_{1}\lambda_{1}+\alpha_{2}\lambda_{2}\right)\right)\\
&\times \delta^{2|8}\left(\tilde{\lambda}_{1}-\alpha_{1}\alpha_{3}\alpha_{5}\alpha_{5'}\tilde{\lambda}_{3}\right)\delta^{2|8}\left(\tilde{\lambda}_{2}-\alpha_{2}\alpha_{3}\alpha_{5}\alpha_{5'}\tilde{\lambda}_{3}\right)
\delta^{2|8}\left(\tilde{\lambda}_{4}-\alpha_{3}\alpha_{4}\tilde{\lambda}_{3}\right)\\&
\times
\delta^{2}\left(\lambda_{3}-\alpha_{3}\left(\alpha_{5}\alpha_{5'}\left(\alpha_{1}\lambda_{1}+\alpha_{2}\lambda_{2}\right)-\alpha_{4}\lambda_{4}\right)\right)
\end{align*}
where we plugged the arguments of the delta functions from the second
line into the remaining delta functions in order to remove their dependence on $\lambda_{5}$ and $\tilde{\lambda}_{5}$. The integrals over $\lambda_{5},\eta_{5}$ can then be trivially
performed against the delta functions in the first line
. For the remaining integral over $\tilde{\lambda}_{5}$, we choose as
integration variables, the second component $\tilde{\lambda}_{5}^2$ so that
the measure $d^2\tilde \lambda_5/GL(1)=\lambda_5^1 d\lambda_5^2$
together with $\alpha_5'$. We
thus get
\[
\int \frac{d \alpha_5'}{(\alpha_5')^3}
\int\frac{d^{2}\tilde{\lambda}_{5}}{GL(1)}\delta^{2}\left(\tilde{\lambda}_{5}-\alpha_{3}\alpha_{5'}\tilde{\lambda}_{3}\right)=1/(\alpha_5')^2
\qquad \alpha_5'=\frac{\tilde \lambda_5^1}{ \alpha_3\tilde \lambda_3^1}\ .
\]
Defining $\alpha_5^{\text{new}}=\alpha_5 \alpha_5'$ (and then dropping
the ``$\text{new}$'', we see that \eqref{eq:bb} finally
reduces to 
\[
\mathcal{A}_{bbs}=\left\langle 12\right\rangle \left\langle 34\right\rangle \int\frac{d(\alpha_{1}\alpha_5\alpha_3)d(\alpha_{2}\alpha_5\alpha_3)d(\alpha_{4}\alpha_3)}{\alpha_{1}^{2}\alpha_{2}^{2}\alpha_{3}^{6}\alpha_{4}^{2}\alpha_{5}^{4}}\delta^{2|8}\left(\tilde{\lambda}_{1}-\alpha_{1}\alpha_5\alpha_{3}\tilde{\lambda}_{3}\right)\delta^{2|8}\left(\tilde{\lambda}_{2}-\alpha_{2}\alpha_5\alpha_{3}\tilde{\lambda}_{3}\right)
\]
\[
\delta^{2|8}\left(\tilde{\lambda}_{4}-\alpha_{3}\alpha_{4}\tilde{\lambda}_{3}\right)\delta^{2}\left(\lambda_{3}-\alpha_{1}\alpha_5\alpha_3\lambda_{1}-\alpha_{2}\alpha_5\alpha_3\lambda_{2}-\alpha_{4}\alpha_3\lambda_{4}\right).
\]
In the next section we will give a simple algorithm which will allow
us to read off this formula directly from the corresponding  graph. 

\begin{figure}
\centering
       \includegraphics{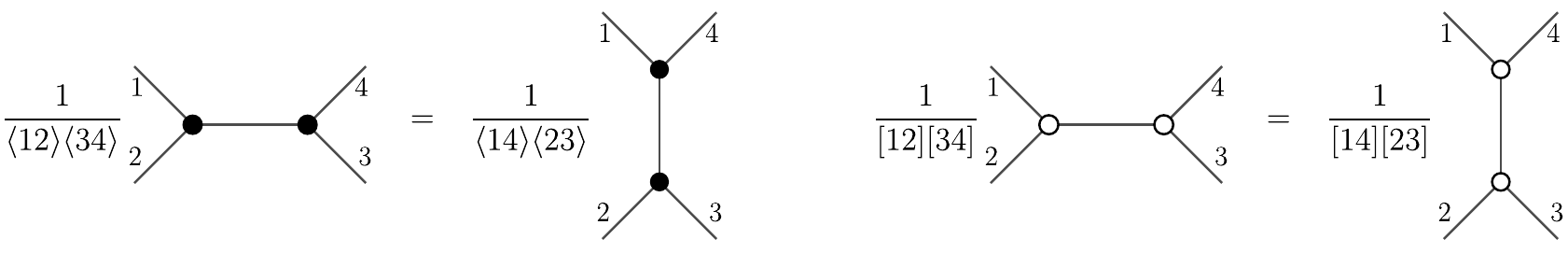}
       \includegraphics{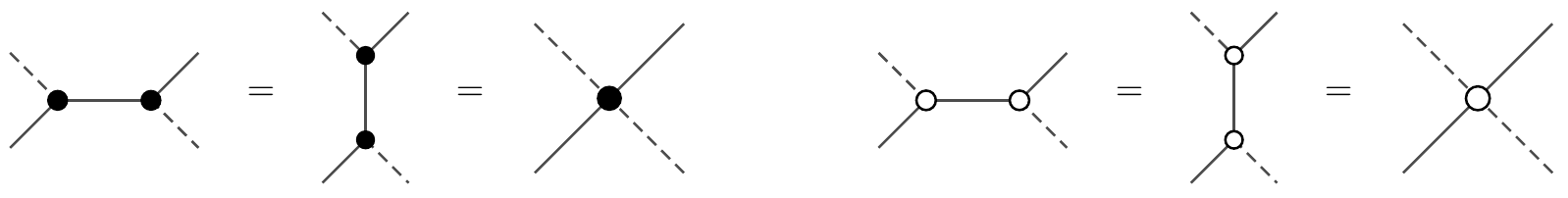}  
    \caption{The merger of like nodes in on-shell diagrams appears
      together with a factor of spinor brackets in
      SUGRA. Alternatively, the merger occurs without a factor if two opposite edges appear with the bridge decoration.} 
    \label{merge1}
    \end{figure}

The equation above can be uplifted to the following  Grassmannian invariant  integral:
\begin{equation}
\mathcal{A}_{bbs}=\left\langle 12\right\rangle \left\langle 34\right\rangle \int\frac{d^{3}C}{(123)^{2}(234)^{2}(341)^{2}(124)^2}\Pi_{\alpha=1}^{3}\delta^{2|8}\left(C_{\alpha}\cdot\tilde{\lambda}\right)\delta^{2}\left(\lambda\cdot C^{\perp}\right)\label{eq:bbg}
\end{equation}
where we recover the previous expression using the  coordinates
\[
C=\left(\begin{array}{cccc}
1 & 0 & -\alpha_{1}\alpha_3\alpha_5 & 0\\
0 & 1 & -\alpha_{2}\alpha_3\alpha_5 & 0\\
0 & 0 & -\alpha_{4}\alpha_3 & 1
\end{array}\right),\,\,\, C^{\perp}=\left(\begin{array}{c}
-\alpha_{1}\alpha_3\alpha_5\\
-\alpha_{2}\alpha_3\alpha_5\\
1\\
-\alpha_{4}\alpha_3
\end{array}\right).
\]
Using a similar analysis, the two-node diagram:
\[  \mathcal{A}_{bbt}=\raisebox{-.5\height}{\includegraphics{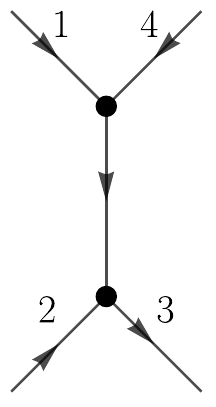}}  
\]
 is given by:
\begin{equation}
\mathcal{A}_{bbt}=\left\langle 14\right\rangle \left\langle 23\right\rangle \int\frac{d^{3}C}{(123)^{2}(234)^{2}(341)^{2}(124)^2}\Pi_{\alpha=1}^{3}\delta^{2|8}\left(C_{\alpha}\cdot\tilde{\lambda}\right)\delta^{2}\left(\lambda\cdot C^{\perp}\right).\label{eq:wwg}
\end{equation}
We see that the expressions in \eqref{eq:bbg} and \eqref{eq:wwg}
are the same up to a prefactor, and analogous relations hold for
two-node diagrams with white vertices. In summary, we have shown how
to glue to like nodes together using edge variables and 
that such diagrams obey the identities
in Figure \ref{merge1}. Furthermore, if two non-adjacent edges are
decorated then the prefactors in the identity are canceled out and it
is possible to define a 4-point vertex by merging together the two
3-point vertices, also depicted in Figure \ref{merge1}.
    
Next, let's consider a two-node diagram with vertices of opposite
color,
\[
  \mathcal{A}_{bw}= \raisebox{-.5\height}{  \includegraphics{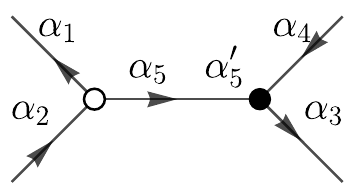} }
  \]
Since we can set an edge variable
to one for each vertex, we will choose $\alpha_{2}=\alpha_{3}=1$.
Using the explicit expressions for 3-point vertices given in the previous subsection,
we find that the diagram is then given by
    \begin{equation}
\mathcal{A}_{bw}=\int\frac{d^{2}\lambda_{5}d^{2}\tilde{\lambda}_{5}d^{8}\eta_{5}}{GL(1)}\mathcal{A}_{3}^{L}\mathcal{A}_{3}^{R}\label{eq:bw}
\end{equation}
where
\[
\mathcal{A}_{3}^{L}=\int\frac{d\alpha_{1}}{\alpha_{1}^{2}}\frac{d\alpha_{5}}{\alpha_{5}^{2}}\left[15\right]\delta^{2|8}\left(\tilde{\lambda}_{2}-\alpha_{1}\tilde{\lambda}_{1}-\alpha_{5}\tilde{\lambda}_{5}\right)\delta^{2}\left(\lambda_{1}-\alpha_{1}\lambda_{2}\right)\delta^{2}\left(\lambda_{5}-\alpha_{5}\lambda_{2}\right)
\]

\[
\mathcal{A}_{3}^{R}=\int\frac{d\alpha_{4}}{\alpha_{4}^{2}}\frac{d\alpha_{5'}}{\alpha_{5'}^{2}}\left\langle 5'4\right\rangle \delta^{2|8}\left(\tilde{\lambda}_{5'}-\alpha_{5'}\tilde{\lambda}_{3}\right)\delta^{2|8}\left(\tilde{\lambda}_{4}-\alpha_{4}\tilde{\lambda}_{3}\right)\delta^{2}\left(\lambda_{3}-\alpha_{5'}\lambda_{5'}-\alpha_{4}\lambda_{4}\right)
\]
and we once again define $\lambda_{5'}=\lambda_{5},\tilde{\lambda}_{5'}=\tilde{\lambda}_{5}$.
Noting that

\[
\left[15\right]=\alpha_{5'}\left[13\right],\,\,\,\left\langle 54\right\rangle =\alpha_{5}\left\langle 24\right\rangle ,
\]
\eqref{eq:bw} can be written as

\[
\mathcal{A}_{bw}=\left[13\right]\left\langle 24\right\rangle \int\frac{d\alpha_{1}}{\alpha_{1}^{2}}\frac{d\alpha_{2}}{\alpha_{4}^{2}}\frac{d\alpha_{5}}{\alpha_{5}}\frac{d\alpha_{5'}}{\alpha_{5'}}\frac{d^{2}\lambda_{5}d^{2}\tilde{\lambda}_{5}d^{8}\eta_{5}}{GL(1)}\delta^{2}\left(\lambda_{5}-\alpha_{5}\lambda_{2}\right)\delta^{2|8}\left(\tilde{\lambda}_{5}-\alpha_{5'}\tilde{\lambda}_{3}\right)
\]
\[
\delta^{2|8}\left(\tilde{\lambda}_{2}-\alpha_{1}\tilde{\lambda}_{1}-\alpha_{5}\alpha_{5'}\tilde{\lambda}_{3}\right)\delta^{2|8}\left(\tilde{\lambda}_{4}-\alpha_{4}\tilde{\lambda}_{3}\right)\delta^{2}\left(\lambda_{1}-\alpha_{1}\lambda_{2}\right)\delta^{2}\left(\lambda_{3}-\alpha_{5}\alpha_{5'}\lambda_{2}-\alpha_{4}\lambda_{4}\right)
\]
where we removed the dependence on $\lambda_{5}$ and $\tilde{\lambda}_{5}$
in the delta functions in second line using delta functions in first
line. The integrals over $\lambda_{5}$ and $\eta_{5}$ are trivial to carry
out and one obtains the constraints $\lambda_{5}=\alpha_{5}\lambda_{2}$,
$\eta_{5}=\alpha_{5'}\eta_{3}$. Using $\mathrm{GL}(1)$ symmetry to set
$\tilde{\lambda}_{5}^{1}=\tilde{\lambda}_{3}^{1},\tilde{\lambda}$ we then obtain
\[
\int\frac{d^{2}\tilde{\lambda}_{5}}{GL(1)}\delta^{2}\left(\tilde{\lambda}_{5}-\alpha_{5'}\tilde{\lambda}_{3}\right)=\delta\left(1-\alpha_{5'}\right)\ .
\] After performing the integral over $\alpha_{5'}$, \eqref{eq:bw}
finally reduces to 
\[
\mathcal{A}_{bw}=\left[13\right]\left\langle 24\right\rangle \int\frac{d\alpha_{1}}{\alpha_{1}^{2}}\frac{d\alpha_{4}}{\alpha_{4}^{2}}\frac{d\alpha_{5}}{\alpha_{5}}\delta^{2|8}\left(\tilde{\lambda}_{2}-\alpha_{1}\tilde{\lambda}_{1}-\alpha_{5}\tilde{\lambda}_{3}\right)\delta^{2|8}\left(\tilde{\lambda}_{4}-\alpha_{4}\tilde{\lambda}_{3}\right)
\]
\[
\times \delta^{2}\left(\lambda_{1}-\alpha_{1}\lambda_{2}\right)\delta^{2}\left(\lambda_{3}-\alpha_{5}\lambda_{2}-\alpha_{4}\lambda_{4}\right),
\]
This can in turn be can be expressed as the residue of a Grassmannian integral as follows:
\begin{equation}
\mathcal{A}_{bw}=\text{Res}_{(12)=0} \int\frac{d^{4}C}{(12)(13)(14)(23)(24)(34)}\frac{\left\langle 24\right\rangle }{(24)}\frac{\left[13\right]}{\left(13\right)^\perp}\Pi_{\alpha=1}^{2}\delta^{2|8}\left(C_{\alpha}\cdot\tilde{\lambda}\right)\Pi_{\beta=1}^{2}\delta^{2}\left(\lambda\cdot C_{\beta}^{\perp}\right),
\label{factorization}
\end{equation}
where the explicit coordinates above correspond to  
\[
C=\left(\begin{array}{cccc}
-\alpha_{1} & 1 & -\alpha_{5} & 0\\
-\alpha & 0 & -\alpha_{4} & 1
\end{array}\right),\,\,\, C^{\perp}=\left(\begin{array}{cc}
1 & 0\\
-\alpha_{1} & -\alpha_{5}\\
0 & 1\\
-\alpha & -\alpha_{4}
\end{array}\right)
\]
and the residue is at $\alpha=0$. 

In summary, we have found that although there are initially two edge
variables associated with a given internal line (one associated to
each end of the line), we can use the $\mathrm{GL}(1)$ symmetry of the on-shell variables of the internal line to set one of the edge variables to one, so that in the end
there is only one edge variable associated with each internal line.
 Moreover, we see the emergence of
Grassmannian structure at the level of two-node diagrams. All of
these features continue to hold for more complicated diagrams.

\subsection{Algorithm} \label{algorithm}

As we have seen from the simple examples in the previous subsection, it is possible to derive Grassmannian integral formulae for higher point on-shell diagrams by combining the 3-point Grassmannians
in \eqref{eq:3ptgrass} and \eqref{eq:3ptmb}. Moreover,
the canonical coordinates -- edge variables -- for these Grassmannians can be 
read off directly from the on-shell diagrams together with a perfect orientation. Scattering amplitudes are then obtained by decorating planar on-shell diagrams with BCFW bridge factors and summing over permutations of the external legs. After doing so, one obtains Grassmannian integral formulae for the scattering amplitudes.

The general algorithm for obtaining a Grassmannian integral formula corresponding to any
on-shell diagram is as follows:
\begin{enumerate}
\item Choose a perfect orientation for the diagram by drawing arrows on each edge such that there are two arrows entering/one arrow leaving every black node and two arrows leaving/one arrow entering  every white node.
  
\item To begin with, label every half-edge with an edge variable $\alpha$ so that there are initially two variables for each internal edge (one associated with each of the two vertices attached to the edge). Then a) set one of the two 
  edge variables on each internal edge to unity, and b) set one of
  the remaining variables associated with each vertex to unity.%
\footnote{The
choices made in a) and b) are arbitrary and final answer should not
depend on this choice. Indeed after a) we are tempted to say there is just a single edge
variable 
for each edge. However in order to implement the intermediate steps of the algorithm
below we need 
to think of it as being associated with one of the two  vertices at the
end of the edge.}
 We
  are thus
  left with 
  $e-v$ independent edge variables.\footnote{From Euler's formula $e-v=f-1$ and as shown in~\cite{ArkaniHamed:2012nw} one can equivalently use face variables. In this context, the edge variables are easier to deal with.}

\item Associate $d\alpha/\alpha^2$ with each edge variable
    leaving a white vertex or entering a black vertex and
    $d\alpha/\alpha^3$ with each edge variable entering a white vertex or
    leaving a black vertex.

  \item For each black vertex associate the bracket $\langle ij \rangle$ where $i$,$j$ are the two edges with ingoing arrows. For each white vertex associate the bracket $[ij]$ where $i$,$j$ are the two edges with outgoing arrows.

  \item All spinor variables (external and internal) are related to
    each other via formulae similar to~\eqref{eq:5}
    and~\eqref{eq:6}:
\begin{align}\label{eq:5b}
  \tilde \lambda_i&=\sum_{\substack{\text{paths}\\ i\rightarrow j}}
  \Big(\prod_{\substack{\text{edges}\\\text{in path}}} \alpha
  \Big)\tilde \lambda_j\ \notag\\
  \lambda_i&=\sum_{\substack{\text{paths}\\ i\leftarrow j}}
  \Big(\prod_{\substack{\text{edges}\\\text{in path}}}\alpha
  \Big)
\lambda_j
\end{align}
Hence, for $\tilde{\lambda}$'s (as well as $\eta$'s) we sum over all paths from edge $i$ to edge $j$, taking the product of all the edge variables encountered along each path, and for $\lambda$'s we consider reverse paths.  If one encounters a closed loop when summing over paths, simply sum the geometric series. 

These relations allow one rewrite the internal spinors in terms of external ones.\footnote {A canonical way to do this
is to simply follow the paths to the end, however one can sometimes obtain simpler expressions by
making more judicious choices, as we will see in the examples in the next section.} Using these relations, write down $\delta^{k\times(2|8)}(C \cdot (\tilde \lambda|\eta))
     \delta^{2(n-k)}(\lambda \cdot C^\perp)$ where $C$ can be
     read off by writing all incoming external $\tilde \lambda$'s in
     terms of outgoing ones and $C^\perp$ can be read off by writing all outgoing external $\lambda$'s in
     terms of ingoing ones.

\item       The above procedure  gives an expression for the on-shell
  diagram as a Grassmannian integral in terms of specific
  coordinates. This can be uplifted to a covariant
  expression by computing the minors of $C$ in terms of edge variables as described in the previous step, and expressing the rest of the integrand in terms of minors whilst ensuring the overall $\mathrm{GL}(k)$ weight is correct (where $k$ is the MHV degree). Note that it is always possible to express the edge variables as monomials of the minors, as was first seen in the context of $\mathcal{N}=4$ SYM \cite{ArkaniHamed:2012nw}. For on-shell diagrams contributing to non-MHV amplitudes, this lift will specify a nontrivial contour in the Grassmannian. We describe this in more detail in the end of section \ref{examples}.   
          \end{enumerate}

Although the above algorithm will work in general, there are often shortcuts one can take to simplify the calculation. Indeed, if an edge of the diagram corresponds to a BCFW bridge, then the spinor brackets associated with the two vertices of this edge will be canceled by the bridge decoration (which has the form $1/ p\cdot p$), leaving only $\alpha$ variables. We can therefore add the following rules to the above algorithm:
\begin{itemize}
\item There is a simple rule for BCFW bridges:
\[    \includegraphics{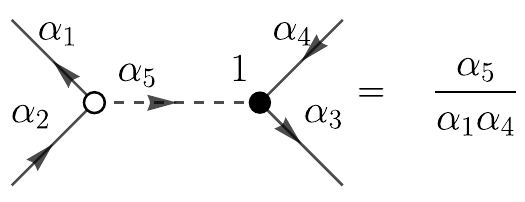} 
\]
  In particular, if $\alpha$ is the edge
  variable of the bridge and all the adjacent edges have trivial edge
  variables, then the bridge contributes $d\alpha/\alpha$.

\item If one uses the planar recursion relation illustrated in Figure~\ref{planar}, one can see that nearly all vertices are attached to bridges. Indeed, for tree-level on-shell diagrams, the only vertices which are not attached to bridges are those directly attached to the $n-2$ unfixed external momenta, so we only need to include spinor brackets for these vertices when implementing step 4 of the algorithm. This observation also has important implications for on-shell diagrams containing bubbles, which should be relevant for loop-level amplitudes. As pointed out in \cite{Herrmann:2016qea}, an undecorated bubble like the one depicted in Figure \ref{bubble} must vanish because the spinor brackets associated with each vertex vanish. On the other hand, if one of the internal lines is decorated then the bubble will not vanish because the decoration precisely cancels out the spinor brackets. Hence, if it is possible to extend BCFW recursion for $\mathcal{N}=8$ supergravity to loop-level, we expect this to be a general feature.
  \begin{figure}
    \centering
    \includegraphics{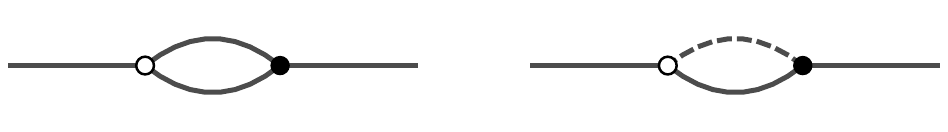}
    \caption{An undecorated bubble diagram (left) vanishes in supergravity, whereas the bridge decoration needed for BCFW renders it finite (right diagram). }
    \label{bubble}
  \end{figure}

\end{itemize}
In the next section, we illustrate this algorithm in a number of examples.

\section{Examples} \label{examples}

In Section \ref{recursion} we described how to recursively compute tree-level amplitudes of $\mathcal{N}=8$ SUGRA  in terms of on-shell diagrams, and in section \ref{grassmannian} we proposed an algorithm for computing the on-shell diagrams in terms of Grassmannian integral formulae. In this section, we will put everything together and illustrate these techniques by computing four and five point amplitudes. In the end of this section, we briefly comment on how these calculations extend to higher-point and in particular non-MHV amplitudes. 

\subsection{Four points}

First we consider the following diagram contributing to the
four-point tree-level  amplitude:
\begin{center}
  \includegraphics{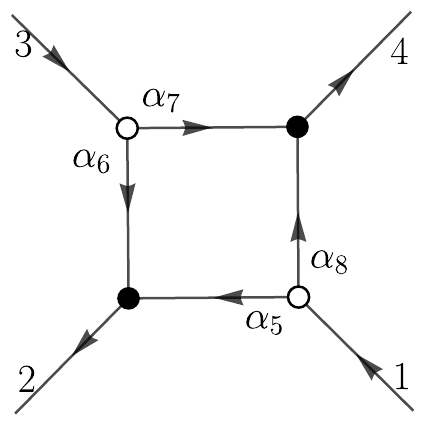}
\end{center}
Here we have already performed the first two steps by orienting and
labeling the diagram. Following steps 3,4 we then have the
expression
\begin{align}
 \int \frac{  d\alpha_5 d\alpha_6 d\alpha_7
  d\alpha_8}{\alpha_5^2\alpha_6^2\alpha_7^2\alpha_8^2
  }\langle67\rangle \langle58\rangle[56][78]\ .\label{eq:7}
\end{align}
We then use the path prescription~\eqref{eq:5b} to rewrite the internal brackets as
external ones
\begin{align}
  \tilde\lambda_5&=\tilde \lambda_2  & \tilde \lambda_6 &=\tilde
                                             \lambda_2 &
  \tilde\lambda_7&=\tilde \lambda_4  & \tilde \lambda_8 &=\tilde
                                             \lambda_4\notag\\
  \lambda_5&= \alpha_5\lambda_1  &  \lambda_6 &=
                                             \alpha_6\lambda_3
 & \lambda_7&= \alpha_7\lambda_3  &  \lambda_8 &=
                                             \alpha_8\lambda_4\ ,
\end{align}
as well as to write all ingoing external $\tilde \lambda$s in terms of
outgoing ones (and vice versa for the $\lambda$'s) yielding the $C$-matrix and $C^\perp$ matrix
\begin{align}
  \label{eq:9}
  \tilde\lambda_1 &=\alpha_5 \lambda_2+\alpha_8\tilde\lambda_4 &
                    \tilde\lambda_3 &=\alpha_7
                                      \lambda_4+\alpha_6\tilde\lambda_2\notag\\
   \lambda_2 &=\alpha_5 \lambda_1+\alpha_6\lambda_3 &
                    \lambda_4 &=\alpha_7
                                      \lambda_3+\alpha_8\lambda_1\ .
\end{align}
Thus after step 5 we have
\begin{align}
   \int \frac{  d\alpha_5 d\alpha_6 d\alpha_7
  d\alpha_8}{\alpha_5\alpha_6\alpha_7\alpha_8
  }\langle13\rangle^2[24]^2\delta^{k(2|8)}(C \cdot (\tilde
  \lambda|\eta)) \delta^{2(n-k)}(\lambda \cdot C^\perp)
\delta^{2\times(2|8)}(C \cdot (\tilde
  \lambda|\eta)) \delta^{2\times2}(\lambda \cdot C^\perp)\
\label{4dln}
\end{align}
with
\begin{align}
C=\left(
  \begin{array}{cccc}
    1&-\alpha_5&0&-\alpha_8\\
    0&-\alpha_6&1&-\alpha_7
  \end{array}
\right)\qquad C^\perp=\left(
  \begin{array}{cccc}
    -\alpha_5&1&-\alpha_6&0\\
    -\alpha_8&0&-\alpha_7&1
  \end{array}
\right)
\ .\label{eq:8}
\end{align}
We thus obtain an expression as an integral over the Grassmannian
with specific coordinates. To uplift this to a covariant expression,
compute all minors of $C$. From~\eqref{eq:7} we have
\begin{align}
  (12)&=-\alpha_6 &(13)&= 1 &(14)&=-\alpha_7\notag\\
 (23)&=-\alpha_5 &(24)&=\alpha_5\alpha_7-\alpha_6\alpha_8
        \ &(34)&=\alpha_8
\end{align}
from which we rewrite~\eqref{eq:8} as
\begin{align}
\mathcal{A}_{4}=\int\frac{d^{2\times4}C}{GL(2)}\frac{\delta^{4|16}\left(C\cdot\tilde{\lambda}|C\cdot\tilde{\eta}\right)\delta^{4}\left(\lambda\cdot C^{\perp}\right)}{\left(12\right)\left(23\right)\left(34\right)(41)}\frac{\left\langle 13\right\rangle ^{2}\left[24\right]^{2}}{\left(13\right)^{4}}.\label{eq:12}
\end{align}
Note that this expression gives the previous expression in the above
coordinates (using that here the measure $d^{2\times4}C/\mathrm{GL}(2) =
d\alpha_5d\alpha_6d\alpha_7d\alpha_8$) and it is invariant under the local  $\mathrm{GL}(2)$ of the
Grassmanian, and so it is the unique Grassman invariant uplift of the
previous expression.%
\footnote{ Under the $\mathrm{GL}(2)$ transformation $C\rightarrow G C$, the
  measure transforms with $\det(G)^4$ and the delta functions with
  $\det(G)^4$, so we need 8 minors in the denominator in order to have $\mathrm{GL}(2)$ invariance, and hence to have a
true Grassmannian integral.}

Before continuing, it is useful to look at what we would get from a different
choice of perfect orientation. Following the steps above for the following
perfect orientation
\begin{center}
  \includegraphics{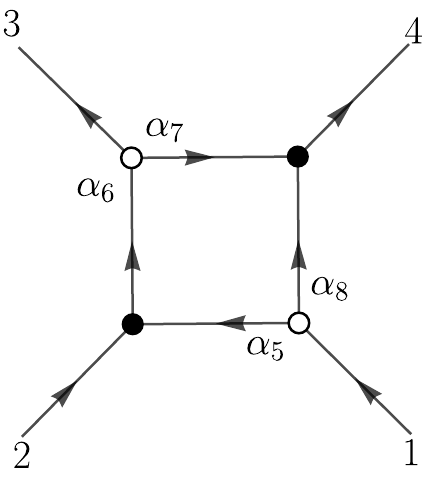}
\end{center}
we obtain the expression 
\begin{align}
\mathcal{A}_{4}=\int\frac{d^{2\times4}C}{GL(2)}\frac{\delta^{4|16}\left(C\cdot\tilde{\lambda}|C\cdot\tilde{\eta}\right)\delta^{4}\left(\lambda\cdot C^{\perp}\right)}{\left(12\right)\left(23\right)\left(34\right)(41)}\frac{\left\langle 12\right\rangle ^{2}\left[34\right]^{2}}{\left(12\right)^{4}}.
\label{4ptu}
\end{align}
This is clearly very similar to the expression we found using the
previous perfect orientation~\eqref{eq:12}, but
for the last factor.   Following similar arguments to those leading
to~\eqref{eq:an} and~\eqref{eq:sq} we have that
\begin{align}
\frac{  \langle13\rangle}{(13)}=\frac{  \langle ij\rangle}{(ij)} \qquad
  \text{and}\qquad \frac{  [24]}{(13)}=\frac{
  [ij]}{(ij)^\perp} \qquad \text{for any } i,j\ ,\label{eq:11}
\end{align}
showing that the two expressions found using different perfect
orientations are in fact equivalent.

Finally, to obtain the 4-point amplitude itself, we multiply by the bridge factor
$\left(\left\langle 12\right\rangle \left[12\right]\right)^{-1}$
and sum over the permutation of legs $3$ and $4$. Using \eqref{eq:11} to choose the last factor in \eqref{4ptu} to be
\begin{align}
  \label{eq:10}
 \frac{ \langle12\rangle^2}{(12)^2}\frac{[12]^2}{(34)^2}\ ,
\end{align}
dividing through by $\left(\left\langle 12\right\rangle \left[12\right]\right)^{-1}$,
and summing over the permutation of legs $3$ and $4$ we obtain
\begin{align}
\label{4ptd}
  \mathcal{M}_{4}&=\int\frac{d^{2\times4}C}{GL(2)}\frac{\delta^{4|16}\left(C\cdot\tilde{\lambda}|C\cdot\tilde{\eta}\right)\delta^{4}\left(\lambda\cdot
  C^{\perp}\right)((13)(24)-(14)(23))}{\prod_{i<j}\left(ij\right)}\frac{\left\langle
  12\right\rangle\left[12\right]}{\left(12\right)^{2}(34)^2}\notag \\
&=\int\frac{d^{2\times4}C}{GL(2)}\frac{\delta^{4|16}\left(C\cdot\tilde{\lambda}|C\cdot\tilde{\eta}\right)\delta^{4}\left(\lambda\cdot
  C^{\perp}\right)}{\prod_{i<j}\left(ij\right)}\frac{\left\langle
  12\right\rangle\left[12\right]}{\left(12\right)(34)}.
\end{align}
To obtain the second line, we used the Plucker identity
$ (13)(24)-(14)(23)=(12)(34)$. Once again, the last factor can be written in many ways, so we write the four-point amplitude more generally as
\[
\mathcal{M}_{4}=\int\frac{d^{2\times4}C}{GL(2)}\frac{\delta^{4|16}\left(C\cdot\tilde{\lambda}|C\cdot\tilde{\eta}\right)\delta^{4}\left(\lambda\cdot C^{\perp}\right)}{\Pi_{i<j}(ij)}\frac{\left\langle kl\right\rangle }{\left(kl\right)}\frac{\left[pq\right]}{\left(p^{\perp}q^{\perp}\right)}
\]
where $k,l,p,q$ are any external legs. 

\begin{figure}[htbp]
\centering
       \includegraphics{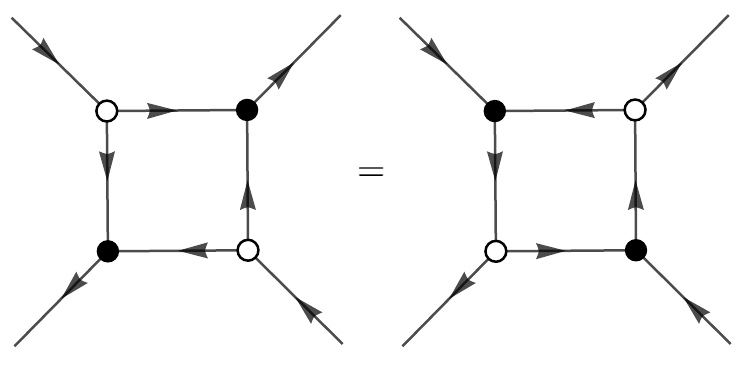}  
    \caption{Square move for on-shell diagrams.}
    \label{square}
    \end{figure}

The Grassmannian integrals in equations \eqref{4ptu} and \eqref{4ptd} are completely localized by the delta functions. In particular, it is not difficult to see that they are solved by 
\[
C=\left(\begin{array}{cccc}
\lambda_{1}^{1} & \lambda_{2}^{1} & \lambda_{3}^{1} & \lambda_{4}^{1}\\
\lambda_{1}^{2} & \lambda_{2}^{2} & \lambda_{3}^{2} & \lambda_{4}^{2}
\end{array}\right),\,\,\, C^{\perp}=\left(\begin{array}{cccc}
\left\langle 34\right\rangle  & 0 & \left\langle 41\right\rangle  & \left\langle 13\right\rangle \\
0 & \left\langle 34\right\rangle  & \left\langle 42\right\rangle  & \left\langle 23\right\rangle 
\end{array}\right).
\]
Evaluating the integrands on these solutions gives the explicit expressions
\begin{equation}
\mathcal{A}_{4}(1,2,3,4)=\left(\frac{\left[34\right]}{\left\langle 12\right\rangle }\right)^{2}\frac{\delta^{4|16}(P)}{\left\langle 12\right\rangle \left\langle 23\right\rangle \left\langle 34\right\rangle \left\langle 41\right\rangle }\label{eq:partial}
\end{equation}
\begin{equation}
\mathcal{M}_{4}=\frac{\left[24\right]}{\left\langle 24\right\rangle \left\langle 13\right\rangle ^{2}}\frac{\delta^{4|16}(P)}{\left\langle 12\right\rangle \left\langle 23\right\rangle \left\langle 34\right\rangle \left\langle 41\right\rangle }.\label{eq:4pttree}
\end{equation}
From the above expressions, one easily sees that the full amplitude $\mathcal{M}_4$ can be obtained from the undecorated partial amplitude $\mathcal{A}_4$ by decorating a bridge and summing over permutations: 
\[
\mathcal{M}_{4}=\frac{1}{\left\langle 12\right\rangle \left[12\right]}\mathcal{A}(1,2,3,4)+3\leftrightarrow4.
\] 
Furthermore, one sees that the on-shell diagrams of $\mathcal{N}=8$ SUGRA are invariant under the square move
depicted in Figure \ref{square}. 
Note that the edge variables of the two diagrams in Figure \ref{square} are nontrivially related, but the integrands are the same because of the invariance of the dlog form of the measure $\Pi_{i=5}^{8}\frac{d\alpha_{i}}{\alpha_{i}}$ appearing in \eqref{4dln}.

\subsection{Five points}

We now move on to the next simplest example, namely five points,  and
apply the algorithm to read off an expression for the planar MHV
on-shell diagram. We choose a  perfect orientation and use the edge
variables according to the diagram
\begin{center}
  \includegraphics{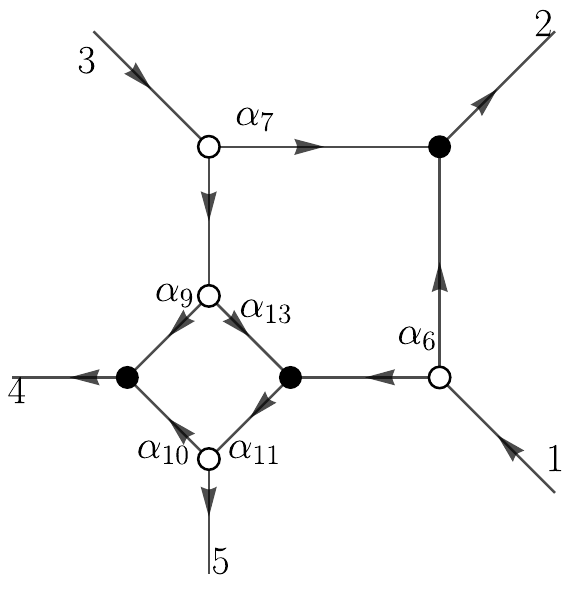}
\end{center}
We have seven internal spinor brackets (one for each vertex) which we rewrite in terms of
external spinors as
\begin{align}
  [6\,12]&=[21] &
                  [5\,10]&=[54]&[13\,9]&=\alpha_{11}[54]&[78]&=[32]\notag\\
\langle67\rangle&=\alpha_6\alpha_7\langle13\rangle&\langle12\,13\rangle&=\alpha_{13}\langle13\rangle&\langle9\,10\rangle&=\alpha_9\alpha_{10}\alpha_{11}\langle31\rangle\ .\label{eq:14}
\end{align}
Inserting these we can thus write down the expression for the diagram as
\begin{align}
  \mathcal{A}_{5}=\int \frac{d\alpha_6d\alpha_7d\alpha_9d\alpha_{10}d\alpha_{11}d\alpha_{13}}{\alpha_6\alpha_7\alpha_9\alpha_{10}\alpha_{13}}   {\delta^{4|16}\left(C\cdot\tilde{\lambda}|C\cdot\tilde{\eta}\right)\delta^{6}\left(\lambda\cdot C^{\perp}\right)}{\left\langle 13\right\rangle ^{3}}{\left[45\right]^{2}\left[12\right]\left[23\right]}.\label{eq:15}
\end{align}
We also read off the $C$ matrix from the diagram 
\begin{align}
  C=\left(
  \begin{array}{ccccc}
    1&-\alpha_6&0&-\alpha_{10}\alpha_{11}&-\alpha_{11}\\
    0&-\alpha_7&1&-\alpha_9-\alpha_{13}\alpha_{11}\alpha_{10}&-\alpha_{13}\alpha_{11}
  \end{array}
\right)
\end{align}
from which we obtain the minors (we only list those which are monomials in the $\alpha$'s)
 \begin{align}
\begin{array}{ccccc}
 (1 2)=-\alpha _7 & (1 3)=1 &  (1 5)=-\alpha _{11} \alpha _{13} & (2
   3)=-\alpha _6 \\
   (3 4)=\alpha _{10} \alpha _{11} & (3 5)=\alpha _{11} & (4
   5)=-\alpha _9 \alpha _{11}. \\
\end{array}
\end{align}
Using the formula for the measure
\begin{align}
\frac{  d^{2\times5}C}{\mathrm{GL(2)}}= \alpha_{11}^3 d\alpha_6d\alpha_7d\alpha_9d\alpha_{10}d\alpha_{11}d\alpha_{13},
\end{align}
we can then uplift the above expression 
directly to the covariant form
\begin{align}
\mathcal{A}_{5}=\int\frac{d^{2\times5}C}{\mathrm{GL(2)}}\frac{\delta^{4|16}\left(C\cdot\tilde{\lambda}|C\cdot\tilde{\eta}\right)\delta^{6}\left(\lambda\cdot C^{\perp}\right)}{\left(12\right)\left(23\right)\left(34\right)(45)(51)}\frac{\left\langle 13\right\rangle ^{3}\left[45\right]^{2}\left[12\right]\left[23\right]}{(13)^{4}}.
\end{align}
(here we need 9 minors in the denominator to get $GL(2)$ invariance).

From the recursion relation in Figure \ref{tree}, we see that the 5-point amplitude
can be obtained by dressing the on-shell diagram with the two BCFW
bridge factors, $1/p_9.p_{11},1/ p_2.p_3$,
 as indicated  by the dashed edges:
\begin{center}
  \includegraphics{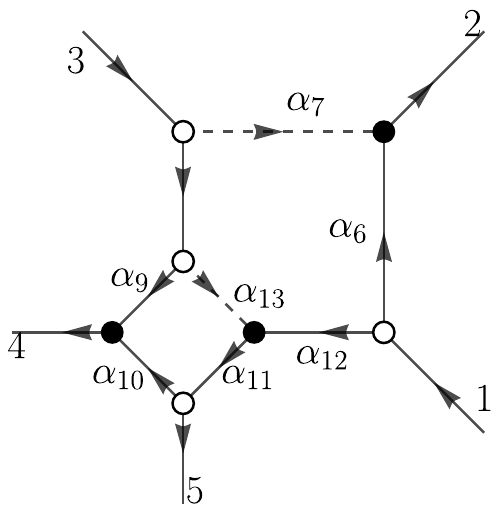}
\end{center}
and summing over permutations of legs $(1,4,5)$. The bridge factors are most naturally incorporated in combination with
the spinor brackets associated with the two vertices attached to each
bridges. So we have 
\begin{align}
 \frac{ [9\,13]\langle12\, 13\rangle}{p_9.p_{11}}&=\frac{\alpha_{13}}{\alpha_9}& \frac{ [78]\langle67\rangle}{p_2.p_{3}}&={\alpha_{7}}
\end{align}
with the remaining three spinor brackets of~(\ref{eq:14}) left untouched. So~(\ref{eq:15}) becomes
\begin{align}
\int
  \frac{d\alpha_6d\alpha_7d\alpha_9d\alpha_{10}d\alpha_{11}d\alpha_{13}}{\alpha_6^2\alpha_7\alpha_9^2\alpha_{10}\alpha_{11}\alpha_{13}}
  {\delta^{4|16}\left(C\cdot\tilde{\lambda}|C\cdot\tilde{\eta}\right)\delta^{6}\left(\lambda\cdot
  C^{\perp}\right)}{[12][45]\langle13\rangle}.\label{eq:15b}
\end{align}
which uplifts to
\begin{align}
\mathcal{A}_{5}^{\mathrm{{decorated}}}(1,2,3,4,5)=\int\frac{d^{2\times5}C}{\mathrm{GL(2)}}\frac{\delta^{4|16}\left(C\cdot\tilde{\lambda}|C\cdot\tilde{\eta}\right)\delta^{6}\left(\lambda\cdot C^{\perp}\right)}{(12)\left(23\right)^{2}\left(34\right)(45)^{2}(51)}\frac{[12][45]\langle13\rangle}{(13)^{2}}.
\end{align}
To obtain the full 5-point amplitude, we must sum the above expression over permutations of $(1,4,5)$. If we first sum over permutations of 4 and 5 and apply a Plucker identity, we obtain
\[
\mathcal{A}_{5}^{\mathrm{{decorated}}}(1,2,3,4,5)+4\leftrightarrow5=[12][45]\int\frac{d^{2\times5}C}{\mathrm{GL(2)}}\frac{\delta^{4|16}\left(C\cdot\tilde{\lambda}|C\cdot\tilde{\eta}\right)\delta^{6}\left(\lambda\cdot C^{\perp}\right)}{(12)\left(23\right)^{2}\left(34\right)(45)(51)(14)(35)}\frac{\langle kl\rangle}{(kl)},
\]
where $k$ and $l$ are any external legs. Summing the above expression over cyclic permutations of $(1,4,5)$ then gives the following expression for the 5-point amplitude:
\[
\mathcal{M}_{5}=\int\frac{d^{2\times5}C}{\mathrm{GL(2)}}\frac{\delta^{4|16}\left(C\cdot\tilde{\lambda}|C\cdot\tilde{\eta}\right)\delta^{6}\left(\lambda\cdot C^{\perp}\right)}{\Pi_{i<j}(ij)}\frac{\left\langle kl\right\rangle }{\left(kl\right)}\frac{N}{\left(23\right)}
\]
where the numerator factor $N$ is 
\[
N=\left(\left[12\right]\left[45\right]\left(13\right)(24)(25)+\left[15\right]\left[24\right]\left(12\right)(25)(34)+\left[25\right]\left[41\right]\left(12\right)(24)(35)\right).
\]
The numerator factor can be simplified by writing it purely in terms of spinor brackets using $(ij)=\frac{\left\langle ij\right\rangle }{\det H}$ and applying momentum conservation and the Schouten identity:
\[
N=\left(23\right)\left(\left[12\right](23)[34](41)-(12)[23](34)[41]\right).
\]
Hence, we obtain the following Grassmannian integral formula for the 5-point amplitude:
\[
\mathcal{M}_{5}=\int\frac{d^{2\times5}C}{\mathrm{GL(2)}}\frac{\delta^{4|16}\left(C\cdot\tilde{\lambda}|C\cdot\tilde{\eta}\right)\delta^{6}\left(\lambda\cdot C^{\perp}\right)}{\Pi_{i<j}(ij)}\frac{\left\langle kl\right\rangle }{\left(kl\right)}\left(\left[12\right](23)[34](41)-(12)[23](34)[41]\right)
\]
where $k,l$ are any external legs. Solving the delta functions and extracting the graviton component of the superamplitude reproduces the five graviton amplitude in the form originally obtained by Berends, Giele, and Kuijf \cite{Berends:1988zp}. 

Let us conclude this section with some general remarks. Using induction, it is not difficult to show that an on-shell diagram contributing to an $n$-point tree-level amplitude will have $n_I = 4(n-3)$ internal edges and $n_V=3n-8$ vertices, regardless of the MHV degree.   From this, it is easy to see that the on-shell diagram will have $n+n_I-n_V=2n-4$ independent edge variables. On the other hand, a tree-level  $n$-point N$^k$MHV amplitude can be expressed as an integral over the Grassmannian $\mathrm{Gr}(n,k)$. Since an element of $\mathrm{Gr}(n,k)$ has $k \times (n-k)$ independent components but an $n$-point on-shell diagram obtained by tree-level BCFW recursion has $(2n-4)$ independent edge variables, this means that when we lift the expression for the on-shell diagram in terms of edge variables into the Grassmannian, this will specify a contour in the Grassmannian of dimension $(k-2)\times (n-k-2)$.

\section{Conclusion}

In this paper, we develop on-shell diagrams for $\mathcal{N}=8$ SUGRA. These are built up from 3-point black and white vertices corresponding to 3-point MHV and anti-MHV amplitudes, respectively. In contrast to $\mathcal{N}=4$ SYM, when computing scattering amplitudes in $\mathcal{N}=8$ SUGRA using BCFW recursion in terms of on-shell diagrams, the BCFW bridge must be decorated and we must sum over permutations of the unshifted external legs. Nevertheless, it is possible to define the recursion in terms of planar on-shell diagrams, implying remarkable new identities for non-planar on-shell diagrams. Moreover, the on-shell diagrams of $\mathcal{N}=8$ SUGRA exhibit equivalence relations analogous to those of $\mathcal{N}=4$ SYM, such as mergers and square moves.

We have also developed an algorithm for computing on-shell diagrams by assigning variables and arrows to the edges in such a way that they form a perfect orientation. This approach is rather appealing because it is very simple and is easy to automate. Furthermore, it leads to a new representation of $\mathcal{N}=8$ SUGRA scattering amplitudes in terms of Grassmannian integral formulae. Other Grassmannian representations were previously deduced using twistor string theory and it would interesting to understand how they are related to our formulae. 

For planar $\mathcal{N}=4$ SYM, the on-shell diagrams were shown to be in one-to-one correspondence with cells of the positive Grassmannian, leading to a new interpretation of scattering amplitudes as the volume of a geometrical object known as the Amplituhedron. We observe hints of similar structure in the undecorated planar on-shell diagrams of $\mathcal{N}=8$ SUGRA from which the amplitudes can be derived after decorating the BCFW bridges and summing over permutations of the external legs. It would therefore be very interesting to explore the existence of an Amplituhedron for $\mathcal{N}=8$ SUGRA. In context of planar $\mathcal{N}=4$ SYM, the Amplituhedron implied the emergence of locality and unitarity from more primitive geometrical principles. If analogous statements can be made for gravitational scattering amplitudes, this may have profound implications for quantum gravity.

Perhaps the most urgent question we face is whether the on-shell diagram formalism we have developed can be extended to loop-level. Whereas dual conformal symmetry provides a canonical definition for the loop integrands of the planar $\mathcal{N}=4$ SYM S-matrix, it is not yet clear how to define a canonical integrand for non-planar (and in particular gravitational) scattering amplitudes, although recent results based on ambitwistor string theory \cite{Geyer:2015bja,Geyer:2015jch} and Q-cuts \cite{Baadsgaard:2015twa} suggest that it is possible to do so. Moreover, BCFW recursion has been used to compute the rational contributions to loop amplitudes in gauge theory \cite{Bern:2005cq,Dunbar:2016aux} and supergravity \cite{Brandhuber:2007up,Dunbar:2016dgg}. 

\begin{figure}
\centering
       \includegraphics{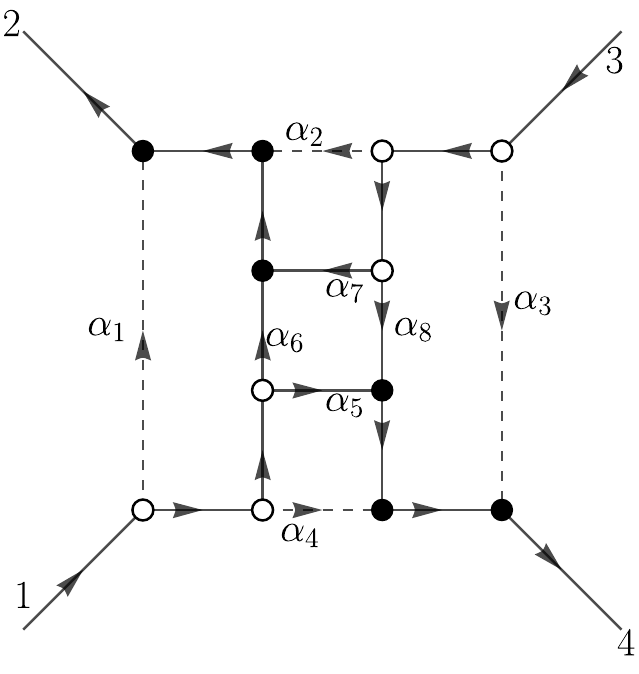}  
\caption{On-shell diagram contributing to the 4-point 1-loop amplitude.}
    \label{1loop}
    \end{figure}

For now, let us simply observe that the one-loop 4-point amplitude of $\mathcal{N}=8$ SUGRA can be obtained from the on-shell diagram depicted in Figure \ref{1loop} after summing over permutations of the external legs. Indeed, using the rules described in section \ref{algorithm}, one immediately finds that this on-shell diagram is given by
\[
\mathcal{A}_{4}^{1}=\int\Pi_{i=1}^{4}\frac{d\alpha_{i}}{\alpha_{i}}\hat{\mathcal{A}}_{4}
\]
where the the dlogs come from the four decorated BCFW bridges, and the integrand simply corresponds to the undecorated planar 4-point on-shell diagram computed in \eqref{eq:partial}:
\[
\hat{\mathcal{A}}_{4}=\frac{\left[\hat{2}\hat{4}\right]^{2}}{\left\langle \hat{1}\hat{3}\right\rangle ^{2}}\frac{\delta^{4|16}(P)}{\left\langle \hat{1}\hat{2}\right\rangle \left\langle \hat{2}\hat{3}\right\rangle \left\langle \hat{3}\hat{4}\right\rangle \left\langle \hat{4}\hat{1}\right\rangle }.
\]
Although the on-shell diagram has eight edge variables, only half of them are independent since the $C$-matrix for this diagram implies the constraints
\[
\lambda_{1}=\left(\alpha_{1}+\alpha_{2}\right)\lambda_{2}+\left(\alpha_{4}+\alpha_{5}\right)\lambda_{4},
\]
\[
\lambda_{3}=\left(\alpha_{2}+\alpha_{7}\right)\lambda_{2}+\left(\alpha_{3}+\alpha_{4}\right)\lambda_{4},
\]
which in turn imply that
\[
\alpha_{5}=\frac{\left\langle 12\right\rangle }{\left\langle 42\right\rangle }-\alpha_{4},\,\,\,\alpha_{6}=\frac{\left\langle 14\right\rangle }{\left\langle 24\right\rangle }-\alpha_{1},\,\,\,\alpha_{7}=\frac{\left\langle 34\right\rangle }{\left\langle 24\right\rangle }-\alpha_{2},\,\,\,\alpha_{8}=\frac{\left\langle 32\right\rangle }{\left\langle 42\right\rangle }-\alpha_{3}.
\]
Hence, the integrand can be written as a function $\left\{ \alpha_{1},\alpha_{2},\alpha_{3},\alpha_{4}\right\}$, as claimed. Noting that $\left[\hat{2}\hat{4}\right]=\left[24\right]$
and $\left\langle \hat{1}\hat{3}\right\rangle =\left\langle 13\right\rangle $ and
dividing by the tree-level 4-point amplitude in \eqref{eq:4pttree} then
gives 
\[
\mathcal{A}_{4}^{1}/\mathcal{M}_{4}=stu\, I_{4}^{1}(s,t)
\]
where $I_{4}^{1}$ is the scalar box integral and we have used the following
identity (which was also used in the context of planar $\mathcal{N}=4$ SYM \cite{ArkaniHamed:2012nw}):
\[
\int\Pi_{i=1}^{4}\frac{d\alpha_{i}}{\alpha_{i}}\frac{\left\langle 12\right\rangle \left\langle 23\right\rangle \left\langle 34\right\rangle \left\langle 41\right\rangle }{\left\langle \hat{1}\hat{2}\right\rangle \left\langle \hat{2}\hat{3}\right\rangle \left\langle \hat{3}\hat{4}\right\rangle \left\langle \hat{4}\hat{1}\right\rangle }=st\, I_{4}^{1}(s,t).
\]
Summing over cyclic permutations of the external legs finally gives the 1-loop
4-point amplitude \cite{Green:1982sw}:
\[
\mathcal{M}_{4}^{1}=stu\mathcal{M}_{4}^{0}\left(I_{4}^{1}(s,t)+I_{4}^{1}(t,u)+I_{4}^{1}(u,s)\right).
\]

Note that the on-shell diagram in Figure \ref{1loop} is just a decorated version of the one corresponding to the 1-loop 4-point amplitude of planar $\mathcal{N}=4$ SYM, which was derived using loop-level BCFW recursion \cite{ArkaniHamed:2012nw}. Our results therefore suggest that a similar recursion relation should exist for $\mathcal{N}=8$ SUGRA, although we leave a detailed derivation for future work. Another interesting feature of the supergravity calculation is the dlog form of the integrand, which is made manifest by the rules described in section \ref{algorithm}. The dlog form was also observed in various other loop amplitudes of $\mathcal{N}=8$ SUGRA \cite{Bern:2014kca}. If it is possible to generalize our on-shell diagram formalism to loop-level, it should make this structure manifest. We also hope that the methods developed in this paper will lead to new techniques for computing non-planar Yang-Mills amplitudes.    

\begin{center}
\textbf{Acknowledgments}
\end{center}

We thank Yvonne Geyer and Lionel Mason for useful conversations. AL is supported by the Royal Society as a Royal Society University Research Fellowship holder, and PH by an STFC Consolidated
Grant ST/L000407/1 and the Marie Curie network
GATIS (gatis.desy.eu) of the European Union’s
Seventh Framework Programme FP7/2007-2013/ under
REA Grant Agreement No. 317089.

\appendix

\section{Grassmannian formulae via the link representation} \label{linka}

In this appendix, we will derive a link representation for the 3-point amplitudes of $\mathcal{N}=8$ SUGRA from which Grassmannian integral formulae can be easily deduced. The link representation for three-point supergravity amplitudes was first considered in \cite{ArkaniHamed:2009si}. 
Consider the 3-point MHV superamplitude:

\[
\mathcal{A}_{3}^{MHV}\left(p_{1},p_{2},p_{3}\right)=\frac{\delta^{4}\left(\lambda\cdot\tilde{\lambda}\right)\delta^{16}\left(\lambda\cdot\eta\right)}{\left\langle 12\right\rangle ^{2}\left\langle 23\right\rangle ^{2}\left\langle 31\right\rangle ^{2}}
\]
where $\lambda\cdot\tilde{\lambda}=\lambda_{1}\tilde{\lambda}_{1}+\lambda_{2}\tilde{\lambda}_{2}+\lambda_{3}\tilde{\lambda}_{3}$
and $\lambda\cdot\eta=\lambda_{1}\eta_{1}+\lambda_{2}\eta_{2}+\lambda_{3}\eta_{3}$.
To obtain a link representation, we first Fourier transform
to twistor space whose coordinates are given by

\[
Z^{A}=\left(\lambda_{\alpha},\mu^{\dot{\alpha}},\tilde{\eta}_{a}\right),\,\,\, W_{A}=\left(\tilde{\mu}^{\alpha},\tilde{\lambda}_{\dot{\alpha}},\eta^{a}\right).
\]
For an N$^{k}$MHV amplitude, one can associate $(k+2)$ legs with $Z$
twistors and the remaining legs with $W$ twistors. Without loss of
generality, let's associate legs 1 and 2 with $Z$ twistors and leg 3
with a $W$ twistor. Then

\[
\mathcal{A}_{3}^{MHV}\left(Z_{1},Z_{2},W_{3}\right)=\int d^{2}\tilde{\lambda}_{1}d^{2}\tilde{\lambda}_{2}d^{2}\lambda_{3}e^{i\left(\mu_{1}\cdot\tilde{\lambda}_{1}+\mu_{2}\cdot\tilde{\lambda}_{2}+\tilde{\mu}_{3}\cdot\lambda_{3}\right)}\mathcal{A}_{3}^{MHV}\left(p_{1},p_{2},p_{3}\right).
\]
Writing the momentum delta function as

\[
\delta^{4}\left(\lambda\cdot\tilde{\lambda}\right)=\int d^{4}xe^{ix\cdot\left(\lambda_{1}\tilde{\lambda}_{1}+\lambda_{2}\tilde{\lambda}_{2}+\lambda_{3}\tilde{\lambda}_{3}\right)},
\]
the integrals over $\tilde{\lambda}_{1}$ and $\tilde{\lambda}_{2}$
give rise to delta functions and we are left with

\begin{equation}
\mathcal{A}_{3}^{MHV}\left(Z_{1},Z_{2},W_{3}\right)=\frac{1}{\left\langle 12\right\rangle ^{2}}\int d^{4}x\delta^{2}\left(\mu_{1}+x\cdot\lambda_{1}\right)\delta^{2}\left(\mu_{2}+x\cdot\lambda_{2}\right)\int d^{2}\lambda_{3}\frac{\delta^{16}\left(\lambda\cdot\eta\right)}{\left\langle 23\right\rangle ^{2}\left\langle 31\right\rangle ^{2}}e^{i\left(\tilde{\mu}_{3}+x\cdot\tilde{\lambda}_{3}\right)\cdot\lambda_{3}}.\label{eq:3pt}
\end{equation}
Next, we express $\lambda_{3}$ as a linear combination of $\lambda_{1}$
and $\lambda_{2}$

\[
\lambda_{3}=c_{13}\lambda_{1}+c_{23}\lambda_{2}
\]
where the coefficients are called link variables We then find that
$d^{2}\lambda_{3}=\left\langle 12\right\rangle dc_{13}dc_{23}$ and

\[
\delta^{16}\left(\lambda\cdot\eta\right)=\left\langle 12\right\rangle ^{4}\delta^{4}\left(\eta_{1}+c_{13}\eta_{3}\right)\delta^{4}\left(\eta_{2}+c_{23}\eta_{3}\right).
\]
Furthermore, on the support of the delta functions in \eqref{eq:3pt},
the argument of the exponential can be expressed in terms of link
variables as follows:

\[
\left(\tilde{\mu}_{3}+x\cdot\tilde{\lambda}_{3}\right)\cdot\lambda_{3}=c_{13}Z_{1}\cdot W_{3}+c_{23}Z_{2}\cdot W_{3}.
\]
Expressing everything in terms of link variables then makes the integral
over $x$ trivial giving a factor of $\left\langle 12\right\rangle ^{-2}$,
leaving us with
\[
\mathcal{A}_{3}^{MHV}\left(Z_{1},Z_{2},W_{3}\right)=\left\langle 12\right\rangle \int\frac{dc_{13}dc_{23}}{c_{13}^{2}c_{23}^{2}}e^{i\left(c_{13}Z_{1}\cdot W_{3}+c_{23}Z_{2}\cdot W_{3}\right)}\delta^{4}\left(\eta_{1}+c_{13}\eta_{3}\right)\delta^{4}\left(\eta_{2}+c_{23}\eta_{3}\right).
\]
Fourier transforming this expression back to momentum space finally gives
\begin{equation}
\mathcal{A}_{3}^{MHV}\left(p_{1},p_{2},p_{3}\right)=\int dc_{13}dc_{23}\frac{\delta^{4|8}\left(C\cdot\lambda|C\cdot\tilde{\eta}\right)\delta^{2}\left(\lambda\cdot C^{\perp}\right)}{\left(12\right)^{2}\left(23\right)^{2}\left(31\right)^{2}}\frac{\left\langle 12\right\rangle }{\left(12\right)}\label{eq:3ptlink}
\end{equation}
where

\[
C=\left(\begin{array}{ccc}
1 & 0 & c_{13}\\
0 & 1 & c_{23}
\end{array}\right),\,\,\, C^{\perp}=\left(\begin{array}{c}
-c_{13}\\
-c_{23}\\
1
\end{array}\right),
\]
and $(ij)$ denotes the minor of columns $i$ and $j$ of $C$. 

Remarkably, \eqref{eq:3ptlink} corresponds to an integral over the Grassmannian
$\mathrm{Gr}(3,2)$. It can be expressed more generally as

\begin{equation}
\mathcal{A}_{3}^{MHV}=\int\frac{d^{2\times3}C}{GL(2)}\frac{\delta^{4|16}\left(C_{\alpha}\cdot\tilde{\lambda}|C_{\alpha}\cdot\tilde{\eta}\right)\delta^{2}\left(\lambda\cdot C^{\perp}\right)}{\left(12\right)^{2}\left(23\right)^{2}\left(31\right)^{2}}\frac{\left\langle ij\right\rangle }{\left(ij\right)}
\label{eq:3ptgrass2a}
\end{equation}
where $i,j$ are any pair of external legs. As explained in Section \ref{3ptg}, the ratio $\frac{\left\langle ij\right\rangle }{\left(ij\right)}$ is the same for any pair of legs $i,j$. Equation \eqref{eq:3ptlink}
corresponds to a particular gauge fixing of the $\mathrm{GL}(2)$ symmetry,
which arose from our choice to associate legs 1 and 2 with $Z$ twistors when deriving the link representation.
Had we made a different choice, we would have obtained a different
gauge fixing of \eqref{eq:3ptgrass2a}. Performing similar manipulations,
we obtain the Grassmannian integral formula for 3-point anti-MHV amplitude in \eqref{eq:3ptmb}.

\end{document}